\documentclass[useAMS,usenatbib]{mn2e}

\title{Galaxy Zoo: Multi-Mergers and the Millennium Simulation}

\author[Darg et al.]{D. W. Darg,$^1$\thanks{Email: ddarg@astro.ox.ac.uk} S. Kaviraj,$^{2,1}$\thanks{Email: skaviraj@astro.ox.ac.uk} C. J. Lintott,$^{1,3}$\thanks{Email: cjl@astro.ox.ac.uk} K. Schawinski,$^{4,5}$\thanks{Einstein Fellow} J. Silk,$^1$ \newauthor S. Lynn,$^{1}$ S. Bamford,$^6$ R. C. Nichol,$^{7}$  \\ \\
$^{1}$ Department of Physics, University of Oxford, Keble Road, Oxford, OX1 3RH, UK\\
$^{2}$ Blackett Laboratory, Imperial College London, London SW7 2AZ \\
$^{3}$ Adler Planetarium, 1300 S. Lake Shore Drive, Chicago, IL 60605, U.S.A.\\
$^{4}$ Department of Physics, Yale University, New Haven, CT 06511, USA\\
$^{5}$ Yale Center for Astronomy and Astrophysics, Yale University, P.O. Box 208121, New Haven, CT 06520, USA\\
$^{6}$ Centre for Astronomy and Particle Theory, University of Nottingham, University Park, Nottingham, NG7 2RD, UK \\
$^{7}$ Institute of Cosmology and Gravitation, University of Portsmouth, Mercantile House, Hampshire Terrace, Portsmouth, PO1 2EG, UK\\
}

\usepackage{graphicx}   
\usepackage{wasysym}	
\usepackage{amsfonts}	%
\usepackage{amssymb}	%
\usepackage{color}
\definecolor{AliceBlue}{rgb}{0.94,0.97,1.00}
\definecolor{AntiqueWhite1}{rgb}{1.00,0.94,0.86}
\definecolor{AntiqueWhite2}{rgb}{0.93,0.87,0.80}
\definecolor{AntiqueWhite3}{rgb}{0.80,0.75,0.69}
\definecolor{AntiqueWhite4}{rgb}{0.55,0.51,0.47}
\definecolor{AntiqueWhite}{rgb}{0.98,0.92,0.84}
\definecolor{BlanchedAlmond}{rgb}{1.00,0.92,0.80}
\definecolor{BlueViolet}{rgb}{0.54,0.17,0.89}
\definecolor{CadetBlue1}{rgb}{0.60,0.96,1.00}
\definecolor{CadetBlue2}{rgb}{0.56,0.90,0.93}
\definecolor{CadetBlue3}{rgb}{0.48,0.77,0.80}
\definecolor{CadetBlue4}{rgb}{0.33,0.53,0.55}
\definecolor{CadetBlue}{rgb}{0.37,0.62,0.63}
\definecolor{CornflowerBlue}{rgb}{0.39,0.58,0.93}
\definecolor{DarkBlue}{rgb}{0.00,0.00,0.55}
\definecolor{DarkCyan}{rgb}{0.00,0.55,0.55}
\definecolor{DarkGoldenrod1}{rgb}{1.00,0.73,0.06}
\definecolor{DarkGoldenrod2}{rgb}{0.93,0.68,0.05}
\definecolor{DarkGoldenrod3}{rgb}{0.80,0.58,0.05}
\definecolor{DarkGoldenrod4}{rgb}{0.55,0.40,0.03}
\definecolor{DarkGoldenrod}{rgb}{0.72,0.53,0.04}
\definecolor{DarkGray}{rgb}{0.66,0.66,0.66}
\definecolor{DarkGreen}{rgb}{0.00,0.39,0.00}
\definecolor{DarkGrey}{rgb}{0.66,0.66,0.66}
\definecolor{DarkKhaki}{rgb}{0.74,0.72,0.42}
\definecolor{DarkMagenta}{rgb}{0.55,0.00,0.55}
\definecolor{DarkOliveGreen1}{rgb}{0.79,1.00,0.44}
\definecolor{DarkOliveGreen2}{rgb}{0.74,0.93,0.41}
\definecolor{DarkOliveGreen3}{rgb}{0.64,0.80,0.35}
\definecolor{DarkOliveGreen4}{rgb}{0.43,0.55,0.24}
\definecolor{DarkOliveGreen}{rgb}{0.33,0.42,0.18}
\definecolor{DarkOrange1}{rgb}{1.00,0.50,0.00}
\definecolor{DarkOrange2}{rgb}{0.93,0.46,0.00}
\definecolor{DarkOrange3}{rgb}{0.80,0.40,0.00}
\definecolor{DarkOrange4}{rgb}{0.55,0.27,0.00}
\definecolor{DarkOrange}{rgb}{1.00,0.55,0.00}
\definecolor{DarkOrchid1}{rgb}{0.75,0.24,1.00}
\definecolor{DarkOrchid2}{rgb}{0.70,0.23,0.93}
\definecolor{DarkOrchid3}{rgb}{0.60,0.20,0.80}
\definecolor{DarkOrchid4}{rgb}{0.41,0.13,0.55}
\definecolor{DarkOrchid}{rgb}{0.60,0.20,0.80}
\definecolor{DarkRed}{rgb}{0.55,0.00,0.00}
\definecolor{DarkSalmon}{rgb}{0.91,0.59,0.48}
\definecolor{DarkSeaGreen1}{rgb}{0.76,1.00,0.76}
\definecolor{DarkSeaGreen2}{rgb}{0.71,0.93,0.71}
\definecolor{DarkSeaGreen3}{rgb}{0.61,0.80,0.61}
\definecolor{DarkSeaGreen4}{rgb}{0.41,0.55,0.41}
\definecolor{DarkSeaGreen}{rgb}{0.56,0.74,0.56}
\definecolor{DarkSlateBlue}{rgb}{0.28,0.24,0.55}
\definecolor{DarkSlateGray1}{rgb}{0.59,1.00,1.00}
\definecolor{DarkSlateGray2}{rgb}{0.55,0.93,0.93}
\definecolor{DarkSlateGray3}{rgb}{0.47,0.80,0.80}
\definecolor{DarkSlateGray4}{rgb}{0.32,0.55,0.55}
\definecolor{DarkSlateGray}{rgb}{0.18,0.31,0.31}
\definecolor{DarkSlateGrey}{rgb}{0.18,0.31,0.31}
\definecolor{DarkTurquoise}{rgb}{0.00,0.81,0.82}
\definecolor{DarkViolet}{rgb}{0.58,0.00,0.83}
\definecolor{DeepPink1}{rgb}{1.00,0.08,0.58}
\definecolor{DeepPink2}{rgb}{0.93,0.07,0.54}
\definecolor{DeepPink3}{rgb}{0.80,0.06,0.46}
\definecolor{DeepPink4}{rgb}{0.55,0.04,0.31}
\definecolor{DeepPink}{rgb}{1.00,0.08,0.58}
\definecolor{DeepSkyBlue1}{rgb}{0.00,0.75,1.00}
\definecolor{DeepSkyBlue2}{rgb}{0.00,0.70,0.93}
\definecolor{DeepSkyBlue3}{rgb}{0.00,0.60,0.80}
\definecolor{DeepSkyBlue4}{rgb}{0.00,0.41,0.55}
\definecolor{DeepSkyBlue}{rgb}{0.00,0.75,1.00}
\definecolor{DimGray}{rgb}{0.41,0.41,0.41}
\definecolor{DimGrey}{rgb}{0.41,0.41,0.41}
\definecolor{DodgerBlue1}{rgb}{0.12,0.56,1.00}
\definecolor{DodgerBlue2}{rgb}{0.11,0.53,0.93}
\definecolor{DodgerBlue3}{rgb}{0.09,0.45,0.80}
\definecolor{DodgerBlue4}{rgb}{0.06,0.31,0.55}
\definecolor{DodgerBlue}{rgb}{0.12,0.56,1.00}
\definecolor{FloralWhite}{rgb}{1.00,0.98,0.94}
\definecolor{ForestGreen}{rgb}{0.13,0.55,0.13}
\definecolor{GhostWhite}{rgb}{0.97,0.97,1.00}
\definecolor{GreenYellow}{rgb}{0.68,1.00,0.18}
\definecolor{HotPink1}{rgb}{1.00,0.43,0.71}
\definecolor{HotPink2}{rgb}{0.93,0.42,0.65}
\definecolor{HotPink3}{rgb}{0.80,0.38,0.56}
\definecolor{HotPink4}{rgb}{0.55,0.23,0.38}
\definecolor{HotPink}{rgb}{1.00,0.41,0.71}
\definecolor{IndianRed1}{rgb}{1.00,0.42,0.42}
\definecolor{IndianRed2}{rgb}{0.93,0.39,0.39}
\definecolor{IndianRed3}{rgb}{0.80,0.33,0.33}
\definecolor{IndianRed4}{rgb}{0.55,0.23,0.23}
\definecolor{IndianRed}{rgb}{0.80,0.36,0.36}
\definecolor{LavenderBlush1}{rgb}{1.00,0.94,0.96}
\definecolor{LavenderBlush2}{rgb}{0.93,0.88,0.90}
\definecolor{LavenderBlush3}{rgb}{0.80,0.76,0.77}
\definecolor{LavenderBlush4}{rgb}{0.55,0.51,0.53}
\definecolor{LavenderBlush}{rgb}{1.00,0.94,0.96}
\definecolor{LawnGreen}{rgb}{0.49,0.99,0.00}
\definecolor{LemonChiffon1}{rgb}{1.00,0.98,0.80}
\definecolor{LemonChiffon2}{rgb}{0.93,0.91,0.75}
\definecolor{LemonChiffon3}{rgb}{0.80,0.79,0.65}
\definecolor{LemonChiffon4}{rgb}{0.55,0.54,0.44}
\definecolor{LemonChiffon}{rgb}{1.00,0.98,0.80}
\definecolor{LightBlue1}{rgb}{0.75,0.94,1.00}
\definecolor{LightBlue2}{rgb}{0.70,0.87,0.93}
\definecolor{LightBlue3}{rgb}{0.60,0.75,0.80}
\definecolor{LightBlue4}{rgb}{0.41,0.51,0.55}
\definecolor{LightBlue}{rgb}{0.68,0.85,0.90}
\definecolor{LightCoral}{rgb}{0.94,0.50,0.50}
\definecolor{LightCyan1}{rgb}{0.88,1.00,1.00}
\definecolor{LightCyan2}{rgb}{0.82,0.93,0.93}
\definecolor{LightCyan3}{rgb}{0.71,0.80,0.80}
\definecolor{LightCyan4}{rgb}{0.48,0.55,0.55}
\definecolor{LightCyan}{rgb}{0.88,1.00,1.00}
\definecolor{LightGoldenrod1}{rgb}{1.00,0.93,0.55}
\definecolor{LightGoldenrod2}{rgb}{0.93,0.86,0.51}
\definecolor{LightGoldenrod3}{rgb}{0.80,0.75,0.44}
\definecolor{LightGoldenrod4}{rgb}{0.55,0.51,0.30}
\definecolor{LightGoldenrodYellow}{rgb}{0.98,0.98,0.82}
\definecolor{LightGoldenrod}{rgb}{0.93,0.87,0.51}
\definecolor{LightGray}{rgb}{0.83,0.83,0.83}
\definecolor{LightGreen}{rgb}{0.56,0.93,0.56}
\definecolor{LightGrey}{rgb}{0.83,0.83,0.83}
\definecolor{LightPink1}{rgb}{1.00,0.68,0.73}
\definecolor{LightPink2}{rgb}{0.93,0.64,0.68}
\definecolor{LightPink3}{rgb}{0.80,0.55,0.58}
\definecolor{LightPink4}{rgb}{0.55,0.37,0.40}
\definecolor{LightPink}{rgb}{1.00,0.71,0.76}
\definecolor{LightSalmon1}{rgb}{1.00,0.63,0.48}
\definecolor{LightSalmon2}{rgb}{0.93,0.58,0.45}
\definecolor{LightSalmon3}{rgb}{0.80,0.51,0.38}
\definecolor{LightSalmon4}{rgb}{0.55,0.34,0.26}
\definecolor{LightSalmon}{rgb}{1.00,0.63,0.48}
\definecolor{LightSeaGreen}{rgb}{0.13,0.70,0.67}
\definecolor{LightSkyBlue1}{rgb}{0.69,0.89,1.00}
\definecolor{LightSkyBlue2}{rgb}{0.64,0.83,0.93}
\definecolor{LightSkyBlue3}{rgb}{0.55,0.71,0.80}
\definecolor{LightSkyBlue4}{rgb}{0.38,0.48,0.55}
\definecolor{LightSkyBlue}{rgb}{0.53,0.81,0.98}
\definecolor{LightSlateBlue}{rgb}{0.52,0.44,1.00}
\definecolor{LightSlateGray}{rgb}{0.47,0.53,0.60}
\definecolor{LightSlateGrey}{rgb}{0.47,0.53,0.60}
\definecolor{LightSteelBlue1}{rgb}{0.79,0.88,1.00}
\definecolor{LightSteelBlue2}{rgb}{0.74,0.82,0.93}
\definecolor{LightSteelBlue3}{rgb}{0.64,0.71,0.80}
\definecolor{LightSteelBlue4}{rgb}{0.43,0.48,0.55}
\definecolor{LightSteelBlue}{rgb}{0.69,0.77,0.87}
\definecolor{LightYellow1}{rgb}{1.00,1.00,0.88}
\definecolor{LightYellow2}{rgb}{0.93,0.93,0.82}
\definecolor{LightYellow3}{rgb}{0.80,0.80,0.71}
\definecolor{LightYellow4}{rgb}{0.55,0.55,0.48}
\definecolor{LightYellow}{rgb}{1.00,1.00,0.88}
\definecolor{LimeGreen}{rgb}{0.20,0.80,0.20}
\definecolor{MediumAquamarine}{rgb}{0.40,0.80,0.67}
\definecolor{MediumBlue}{rgb}{0.00,0.00,0.80}
\definecolor{MediumOrchid1}{rgb}{0.88,0.40,1.00}
\definecolor{MediumOrchid2}{rgb}{0.82,0.37,0.93}
\definecolor{MediumOrchid3}{rgb}{0.71,0.32,0.80}
\definecolor{MediumOrchid4}{rgb}{0.48,0.22,0.55}
\definecolor{MediumOrchid}{rgb}{0.73,0.33,0.83}
\definecolor{MediumPurple1}{rgb}{0.67,0.51,1.00}
\definecolor{MediumPurple2}{rgb}{0.62,0.47,0.93}
\definecolor{MediumPurple3}{rgb}{0.54,0.41,0.80}
\definecolor{MediumPurple4}{rgb}{0.36,0.28,0.55}
\definecolor{MediumPurple}{rgb}{0.58,0.44,0.86}
\definecolor{MediumSeaGreen}{rgb}{0.24,0.70,0.44}
\definecolor{MediumSlateBlue}{rgb}{0.48,0.41,0.93}
\definecolor{MediumSpringGreen}{rgb}{0.00,0.98,0.60}
\definecolor{MediumTurquoise}{rgb}{0.28,0.82,0.80}
\definecolor{MediumVioletRed}{rgb}{0.78,0.08,0.52}
\definecolor{MidnightBlue}{rgb}{0.10,0.10,0.44}
\definecolor{MintCream}{rgb}{0.96,1.00,0.98}
\definecolor{MistyRose1}{rgb}{1.00,0.89,0.88}
\definecolor{MistyRose2}{rgb}{0.93,0.84,0.82}
\definecolor{MistyRose3}{rgb}{0.80,0.72,0.71}
\definecolor{MistyRose4}{rgb}{0.55,0.49,0.48}
\definecolor{MistyRose}{rgb}{1.00,0.89,0.88}
\definecolor{NavajoWhite1}{rgb}{1.00,0.87,0.68}
\definecolor{NavajoWhite2}{rgb}{0.93,0.81,0.63}
\definecolor{NavajoWhite3}{rgb}{0.80,0.70,0.55}
\definecolor{NavajoWhite4}{rgb}{0.55,0.47,0.37}
\definecolor{NavajoWhite}{rgb}{1.00,0.87,0.68}
\definecolor{NavyBlue}{rgb}{0.00,0.00,0.50}
\definecolor{OldLace}{rgb}{0.99,0.96,0.90}
\definecolor{OliveDrab1}{rgb}{0.75,1.00,0.24}
\definecolor{OliveDrab2}{rgb}{0.70,0.93,0.23}
\definecolor{OliveDrab3}{rgb}{0.60,0.80,0.20}
\definecolor{OliveDrab4}{rgb}{0.41,0.55,0.13}
\definecolor{OliveDrab}{rgb}{0.42,0.56,0.14}
\definecolor{OrangeRed1}{rgb}{1.00,0.27,0.00}
\definecolor{OrangeRed2}{rgb}{0.93,0.25,0.00}
\definecolor{OrangeRed3}{rgb}{0.80,0.22,0.00}
\definecolor{OrangeRed4}{rgb}{0.55,0.15,0.00}
\definecolor{OrangeRed}{rgb}{1.00,0.27,0.00}
\definecolor{PaleGoldenrod}{rgb}{0.93,0.91,0.67}
\definecolor{PaleGreen1}{rgb}{0.60,1.00,0.60}
\definecolor{PaleGreen2}{rgb}{0.56,0.93,0.56}
\definecolor{PaleGreen3}{rgb}{0.49,0.80,0.49}
\definecolor{PaleGreen4}{rgb}{0.33,0.55,0.33}
\definecolor{PaleGreen}{rgb}{0.60,0.98,0.60}
\definecolor{PaleTurquoise1}{rgb}{0.73,1.00,1.00}
\definecolor{PaleTurquoise2}{rgb}{0.68,0.93,0.93}
\definecolor{PaleTurquoise3}{rgb}{0.59,0.80,0.80}
\definecolor{PaleTurquoise4}{rgb}{0.40,0.55,0.55}
\definecolor{PaleTurquoise}{rgb}{0.69,0.93,0.93}
\definecolor{PaleVioletRed1}{rgb}{1.00,0.51,0.67}
\definecolor{PaleVioletRed2}{rgb}{0.93,0.47,0.62}
\definecolor{PaleVioletRed3}{rgb}{0.80,0.41,0.54}
\definecolor{PaleVioletRed4}{rgb}{0.55,0.28,0.36}
\definecolor{PaleVioletRed}{rgb}{0.86,0.44,0.58}
\definecolor{PapayaWhip}{rgb}{1.00,0.94,0.84}
\definecolor{PeachPuff1}{rgb}{1.00,0.85,0.73}
\definecolor{PeachPuff2}{rgb}{0.93,0.80,0.68}
\definecolor{PeachPuff3}{rgb}{0.80,0.69,0.58}
\definecolor{PeachPuff4}{rgb}{0.55,0.47,0.40}
\definecolor{PeachPuff}{rgb}{1.00,0.85,0.73}
\definecolor{PowderBlue}{rgb}{0.69,0.88,0.90}
\definecolor{RosyBrown1}{rgb}{1.00,0.76,0.76}
\definecolor{RosyBrown2}{rgb}{0.93,0.71,0.71}
\definecolor{RosyBrown3}{rgb}{0.80,0.61,0.61}
\definecolor{RosyBrown4}{rgb}{0.55,0.41,0.41}
\definecolor{RosyBrown}{rgb}{0.74,0.56,0.56}
\definecolor{RoyalBlue1}{rgb}{0.28,0.46,1.00}
\definecolor{RoyalBlue2}{rgb}{0.26,0.43,0.93}
\definecolor{RoyalBlue3}{rgb}{0.23,0.37,0.80}
\definecolor{RoyalBlue4}{rgb}{0.15,0.25,0.55}
\definecolor{RoyalBlue}{rgb}{0.25,0.41,0.88}
\definecolor{SaddleBrown}{rgb}{0.55,0.27,0.07}
\definecolor{SandyBrown}{rgb}{0.96,0.64,0.38}
\definecolor{SeaGreen1}{rgb}{0.33,1.00,0.62}
\definecolor{SeaGreen2}{rgb}{0.31,0.93,0.58}
\definecolor{SeaGreen3}{rgb}{0.26,0.80,0.50}
\definecolor{SeaGreen4}{rgb}{0.18,0.55,0.34}
\definecolor{SeaGreen}{rgb}{0.18,0.55,0.34}
\definecolor{SkyBlue1}{rgb}{0.53,0.81,1.00}
\definecolor{SkyBlue2}{rgb}{0.49,0.75,0.93}
\definecolor{SkyBlue3}{rgb}{0.42,0.65,0.80}
\definecolor{SkyBlue4}{rgb}{0.29,0.44,0.55}
\definecolor{SkyBlue}{rgb}{0.53,0.81,0.92}
\definecolor{SlateBlue1}{rgb}{0.51,0.44,1.00}
\definecolor{SlateBlue2}{rgb}{0.48,0.40,0.93}
\definecolor{SlateBlue3}{rgb}{0.41,0.35,0.80}
\definecolor{SlateBlue4}{rgb}{0.28,0.24,0.55}
\definecolor{SlateBlue}{rgb}{0.42,0.35,0.80}
\definecolor{SlateGray1}{rgb}{0.78,0.89,1.00}
\definecolor{SlateGray2}{rgb}{0.73,0.83,0.93}
\definecolor{SlateGray3}{rgb}{0.62,0.71,0.80}
\definecolor{SlateGray4}{rgb}{0.42,0.48,0.55}
\definecolor{SlateGray}{rgb}{0.44,0.50,0.56}
\definecolor{SlateGrey}{rgb}{0.44,0.50,0.56}
\definecolor{SpringGreen1}{rgb}{0.00,1.00,0.50}
\definecolor{SpringGreen2}{rgb}{0.00,0.93,0.46}
\definecolor{SpringGreen3}{rgb}{0.00,0.80,0.40}
\definecolor{SpringGreen4}{rgb}{0.00,0.55,0.27}
\definecolor{SpringGreen}{rgb}{0.00,1.00,0.50}
\definecolor{SteelBlue1}{rgb}{0.39,0.72,1.00}
\definecolor{SteelBlue2}{rgb}{0.36,0.67,0.93}
\definecolor{SteelBlue3}{rgb}{0.31,0.58,0.80}
\definecolor{SteelBlue4}{rgb}{0.21,0.39,0.55}
\definecolor{SteelBlue}{rgb}{0.27,0.51,0.71}
\definecolor{VioletRed1}{rgb}{1.00,0.24,0.59}
\definecolor{VioletRed2}{rgb}{0.93,0.23,0.55}
\definecolor{VioletRed3}{rgb}{0.80,0.20,0.47}
\definecolor{VioletRed4}{rgb}{0.55,0.13,0.32}
\definecolor{VioletRed}{rgb}{0.82,0.13,0.56}
\definecolor{WhiteSmoke}{rgb}{0.96,0.96,0.96}
\definecolor{YellowGreen}{rgb}{0.60,0.80,0.20}
\definecolor{aliceblue}{rgb}{0.94,0.97,1.00}
\definecolor{antiquewhite}{rgb}{0.98,0.92,0.84}
\definecolor{aquamarine1}{rgb}{0.50,1.00,0.83}
\definecolor{aquamarine2}{rgb}{0.46,0.93,0.78}
\definecolor{aquamarine3}{rgb}{0.40,0.80,0.67}
\definecolor{aquamarine4}{rgb}{0.27,0.55,0.45}
\definecolor{aquamarine}{rgb}{0.50,1.00,0.83}
\definecolor{azure1}{rgb}{0.94,1.00,1.00}
\definecolor{azure2}{rgb}{0.88,0.93,0.93}
\definecolor{azure3}{rgb}{0.76,0.80,0.80}
\definecolor{azure4}{rgb}{0.51,0.55,0.55}
\definecolor{azure}{rgb}{0.94,1.00,1.00}
\definecolor{beige}{rgb}{0.96,0.96,0.86}
\definecolor{bisque1}{rgb}{1.00,0.89,0.77}
\definecolor{bisque2}{rgb}{0.93,0.84,0.72}
\definecolor{bisque3}{rgb}{0.80,0.72,0.62}
\definecolor{bisque4}{rgb}{0.55,0.49,0.42}
\definecolor{bisque}{rgb}{1.00,0.89,0.77}
\definecolor{black}{rgb}{0.00,0.00,0.00}
\definecolor{blanchedalmond}{rgb}{1.00,0.92,0.80}
\definecolor{blue1}{rgb}{0.00,0.00,1.00}
\definecolor{blue2}{rgb}{0.00,0.00,0.93}
\definecolor{blue3}{rgb}{0.00,0.00,0.80}
\definecolor{blue4}{rgb}{0.00,0.00,0.55}
\definecolor{blueviolet}{rgb}{0.54,0.17,0.89}
\definecolor{blue}{rgb}{0.00,0.00,1.00}
\definecolor{brown1}{rgb}{1.00,0.25,0.25}
\definecolor{brown2}{rgb}{0.93,0.23,0.23}
\definecolor{brown3}{rgb}{0.80,0.20,0.20}
\definecolor{brown4}{rgb}{0.55,0.14,0.14}
\definecolor{brown}{rgb}{0.65,0.16,0.16}
\definecolor{burlywood1}{rgb}{1.00,0.83,0.61}
\definecolor{burlywood2}{rgb}{0.93,0.77,0.57}
\definecolor{burlywood3}{rgb}{0.80,0.67,0.49}
\definecolor{burlywood4}{rgb}{0.55,0.45,0.33}
\definecolor{burlywood}{rgb}{0.87,0.72,0.53}
\definecolor{cadetblue}{rgb}{0.37,0.62,0.63}
\definecolor{chartreuse1}{rgb}{0.50,1.00,0.00}
\definecolor{chartreuse2}{rgb}{0.46,0.93,0.00}
\definecolor{chartreuse3}{rgb}{0.40,0.80,0.00}
\definecolor{chartreuse4}{rgb}{0.27,0.55,0.00}
\definecolor{chartreuse}{rgb}{0.50,1.00,0.00}
\definecolor{chocolate1}{rgb}{1.00,0.50,0.14}
\definecolor{chocolate2}{rgb}{0.93,0.46,0.13}
\definecolor{chocolate3}{rgb}{0.80,0.40,0.11}
\definecolor{chocolate4}{rgb}{0.55,0.27,0.07}
\definecolor{chocolate}{rgb}{0.82,0.41,0.12}
\definecolor{coral1}{rgb}{1.00,0.45,0.34}
\definecolor{coral2}{rgb}{0.93,0.42,0.31}
\definecolor{coral3}{rgb}{0.80,0.36,0.27}
\definecolor{coral4}{rgb}{0.55,0.24,0.18}
\definecolor{coral}{rgb}{1.00,0.50,0.31}
\definecolor{cornflowerblue}{rgb}{0.39,0.58,0.93}
\definecolor{cornsilk1}{rgb}{1.00,0.97,0.86}
\definecolor{cornsilk2}{rgb}{0.93,0.91,0.80}
\definecolor{cornsilk3}{rgb}{0.80,0.78,0.69}
\definecolor{cornsilk4}{rgb}{0.55,0.53,0.47}
\definecolor{cornsilk}{rgb}{1.00,0.97,0.86}
\definecolor{cyan1}{rgb}{0.00,1.00,1.00}
\definecolor{cyan2}{rgb}{0.00,0.93,0.93}
\definecolor{cyan3}{rgb}{0.00,0.80,0.80}
\definecolor{cyan4}{rgb}{0.00,0.55,0.55}
\definecolor{cyan}{rgb}{0.00,1.00,1.00}
\definecolor{darkblue}{rgb}{0.00,0.00,0.55}
\definecolor{darkcyan}{rgb}{0.00,0.55,0.55}
\definecolor{darkgoldenrod}{rgb}{0.72,0.53,0.04}
\definecolor{darkgray}{rgb}{0.66,0.66,0.66}
\definecolor{darkgreen}{rgb}{0.00,0.39,0.00}
\definecolor{darkgrey}{rgb}{0.66,0.66,0.66}
\definecolor{darkkhaki}{rgb}{0.74,0.72,0.42}
\definecolor{darkmagenta}{rgb}{0.55,0.00,0.55}
\definecolor{darkolive}{rgb}{0.33,0.42,0.18}
\definecolor{darkorange}{rgb}{1.00,0.55,0.00}
\definecolor{darkorchid}{rgb}{0.60,0.20,0.80}
\definecolor{darkred}{rgb}{0.55,0.00,0.00}
\definecolor{darksalmon}{rgb}{0.91,0.59,0.48}
\definecolor{darksea}{rgb}{0.56,0.74,0.56}
\definecolor{darkslate}{rgb}{0.18,0.31,0.31}
\definecolor{darkslate}{rgb}{0.18,0.31,0.31}
\definecolor{darkslate}{rgb}{0.28,0.24,0.55}
\definecolor{darkturquoise}{rgb}{0.00,0.81,0.82}
\definecolor{darkviolet}{rgb}{0.58,0.00,0.83}
\definecolor{deeppink}{rgb}{1.00,0.08,0.58}
\definecolor{deepsky}{rgb}{0.00,0.75,1.00}
\definecolor{dimgray}{rgb}{0.41,0.41,0.41}
\definecolor{dimgrey}{rgb}{0.41,0.41,0.41}
\definecolor{dodgerblue}{rgb}{0.12,0.56,1.00}
\definecolor{firebrick1}{rgb}{1.00,0.19,0.19}
\definecolor{firebrick2}{rgb}{0.93,0.17,0.17}
\definecolor{firebrick3}{rgb}{0.80,0.15,0.15}
\definecolor{firebrick4}{rgb}{0.55,0.10,0.10}
\definecolor{firebrick}{rgb}{0.70,0.13,0.13}
\definecolor{floralwhite}{rgb}{1.00,0.98,0.94}
\definecolor{forestgreen}{rgb}{0.13,0.55,0.13}
\definecolor{gainsboro}{rgb}{0.86,0.86,0.86}
\definecolor{ghostwhite}{rgb}{0.97,0.97,1.00}
\definecolor{gold1}{rgb}{1.00,0.84,0.00}
\definecolor{gold2}{rgb}{0.93,0.79,0.00}
\definecolor{gold3}{rgb}{0.80,0.68,0.00}
\definecolor{gold4}{rgb}{0.55,0.46,0.00}
\definecolor{goldenrod1}{rgb}{1.00,0.76,0.15}
\definecolor{goldenrod2}{rgb}{0.93,0.71,0.13}
\definecolor{goldenrod3}{rgb}{0.80,0.61,0.11}
\definecolor{goldenrod4}{rgb}{0.55,0.41,0.08}
\definecolor{goldenrod}{rgb}{0.85,0.65,0.13}
\definecolor{gold}{rgb}{1.00,0.84,0.00}
\definecolor{gray0}{rgb}{0.00,0.00,0.00}
\definecolor{gray100}{rgb}{1.00,1.00,1.00}
\definecolor{gray10}{rgb}{0.10,0.10,0.10}
\definecolor{gray11}{rgb}{0.11,0.11,0.11}
\definecolor{gray12}{rgb}{0.12,0.12,0.12}
\definecolor{gray13}{rgb}{0.13,0.13,0.13}
\definecolor{gray14}{rgb}{0.14,0.14,0.14}
\definecolor{gray15}{rgb}{0.15,0.15,0.15}
\definecolor{gray16}{rgb}{0.16,0.16,0.16}
\definecolor{gray17}{rgb}{0.17,0.17,0.17}
\definecolor{gray18}{rgb}{0.18,0.18,0.18}
\definecolor{gray19}{rgb}{0.19,0.19,0.19}
\definecolor{gray1}{rgb}{0.01,0.01,0.01}
\definecolor{gray20}{rgb}{0.20,0.20,0.20}
\definecolor{gray21}{rgb}{0.21,0.21,0.21}
\definecolor{gray22}{rgb}{0.22,0.22,0.22}
\definecolor{gray23}{rgb}{0.23,0.23,0.23}
\definecolor{gray24}{rgb}{0.24,0.24,0.24}
\definecolor{gray25}{rgb}{0.25,0.25,0.25}
\definecolor{gray26}{rgb}{0.26,0.26,0.26}
\definecolor{gray27}{rgb}{0.27,0.27,0.27}
\definecolor{gray28}{rgb}{0.28,0.28,0.28}
\definecolor{gray29}{rgb}{0.29,0.29,0.29}
\definecolor{gray2}{rgb}{0.02,0.02,0.02}
\definecolor{gray30}{rgb}{0.30,0.30,0.30}
\definecolor{gray31}{rgb}{0.31,0.31,0.31}
\definecolor{gray32}{rgb}{0.32,0.32,0.32}
\definecolor{gray33}{rgb}{0.33,0.33,0.33}
\definecolor{gray34}{rgb}{0.34,0.34,0.34}
\definecolor{gray35}{rgb}{0.35,0.35,0.35}
\definecolor{gray36}{rgb}{0.36,0.36,0.36}
\definecolor{gray37}{rgb}{0.37,0.37,0.37}
\definecolor{gray38}{rgb}{0.38,0.38,0.38}
\definecolor{gray39}{rgb}{0.39,0.39,0.39}
\definecolor{gray3}{rgb}{0.03,0.03,0.03}
\definecolor{gray40}{rgb}{0.40,0.40,0.40}
\definecolor{gray41}{rgb}{0.41,0.41,0.41}
\definecolor{gray42}{rgb}{0.42,0.42,0.42}
\definecolor{gray43}{rgb}{0.43,0.43,0.43}
\definecolor{gray44}{rgb}{0.44,0.44,0.44}
\definecolor{gray45}{rgb}{0.45,0.45,0.45}
\definecolor{gray46}{rgb}{0.46,0.46,0.46}
\definecolor{gray47}{rgb}{0.47,0.47,0.47}
\definecolor{gray48}{rgb}{0.48,0.48,0.48}
\definecolor{gray49}{rgb}{0.49,0.49,0.49}
\definecolor{gray4}{rgb}{0.04,0.04,0.04}
\definecolor{gray50}{rgb}{0.50,0.50,0.50}
\definecolor{gray51}{rgb}{0.51,0.51,0.51}
\definecolor{gray52}{rgb}{0.52,0.52,0.52}
\definecolor{gray53}{rgb}{0.53,0.53,0.53}
\definecolor{gray54}{rgb}{0.54,0.54,0.54}
\definecolor{gray55}{rgb}{0.55,0.55,0.55}
\definecolor{gray56}{rgb}{0.56,0.56,0.56}
\definecolor{gray57}{rgb}{0.57,0.57,0.57}
\definecolor{gray58}{rgb}{0.58,0.58,0.58}
\definecolor{gray59}{rgb}{0.59,0.59,0.59}
\definecolor{gray5}{rgb}{0.05,0.05,0.05}
\definecolor{gray60}{rgb}{0.60,0.60,0.60}
\definecolor{gray61}{rgb}{0.61,0.61,0.61}
\definecolor{gray62}{rgb}{0.62,0.62,0.62}
\definecolor{gray63}{rgb}{0.63,0.63,0.63}
\definecolor{gray64}{rgb}{0.64,0.64,0.64}
\definecolor{gray65}{rgb}{0.65,0.65,0.65}
\definecolor{gray66}{rgb}{0.66,0.66,0.66}
\definecolor{gray67}{rgb}{0.67,0.67,0.67}
\definecolor{gray68}{rgb}{0.68,0.68,0.68}
\definecolor{gray69}{rgb}{0.69,0.69,0.69}
\definecolor{gray6}{rgb}{0.06,0.06,0.06}
\definecolor{gray70}{rgb}{0.70,0.70,0.70}
\definecolor{gray71}{rgb}{0.71,0.71,0.71}
\definecolor{gray72}{rgb}{0.72,0.72,0.72}
\definecolor{gray73}{rgb}{0.73,0.73,0.73}
\definecolor{gray74}{rgb}{0.74,0.74,0.74}
\definecolor{gray75}{rgb}{0.75,0.75,0.75}
\definecolor{gray76}{rgb}{0.76,0.76,0.76}
\definecolor{gray77}{rgb}{0.77,0.77,0.77}
\definecolor{gray78}{rgb}{0.78,0.78,0.78}
\definecolor{gray79}{rgb}{0.79,0.79,0.79}
\definecolor{gray7}{rgb}{0.07,0.07,0.07}
\definecolor{gray80}{rgb}{0.80,0.80,0.80}
\definecolor{gray81}{rgb}{0.81,0.81,0.81}
\definecolor{gray82}{rgb}{0.82,0.82,0.82}
\definecolor{gray83}{rgb}{0.83,0.83,0.83}
\definecolor{gray84}{rgb}{0.84,0.84,0.84}
\definecolor{gray85}{rgb}{0.85,0.85,0.85}
\definecolor{gray86}{rgb}{0.86,0.86,0.86}
\definecolor{gray87}{rgb}{0.87,0.87,0.87}
\definecolor{gray88}{rgb}{0.88,0.88,0.88}
\definecolor{gray89}{rgb}{0.89,0.89,0.89}
\definecolor{gray8}{rgb}{0.08,0.08,0.08}
\definecolor{gray90}{rgb}{0.90,0.90,0.90}
\definecolor{gray91}{rgb}{0.91,0.91,0.91}
\definecolor{gray92}{rgb}{0.92,0.92,0.92}
\definecolor{gray93}{rgb}{0.93,0.93,0.93}
\definecolor{gray94}{rgb}{0.94,0.94,0.94}
\definecolor{gray95}{rgb}{0.95,0.95,0.95}
\definecolor{gray96}{rgb}{0.96,0.96,0.96}
\definecolor{gray97}{rgb}{0.97,0.97,0.97}
\definecolor{gray98}{rgb}{0.98,0.98,0.98}
\definecolor{gray99}{rgb}{0.99,0.99,0.99}
\definecolor{gray9}{rgb}{0.09,0.09,0.09}
\definecolor{gray}{rgb}{0.75,0.75,0.75}
\definecolor{green1}{rgb}{0.00,1.00,0.00}
\definecolor{green2}{rgb}{0.00,0.93,0.00}
\definecolor{green3}{rgb}{0.00,0.80,0.00}
\definecolor{green4}{rgb}{0.00,0.55,0.00}
\definecolor{greenyellow}{rgb}{0.68,1.00,0.18}
\definecolor{green}{rgb}{0.00,1.00,0.00}
\definecolor{grey0}{rgb}{0.00,0.00,0.00}
\definecolor{grey100}{rgb}{1.00,1.00,1.00}
\definecolor{grey10}{rgb}{0.10,0.10,0.10}
\definecolor{grey11}{rgb}{0.11,0.11,0.11}
\definecolor{grey12}{rgb}{0.12,0.12,0.12}
\definecolor{grey13}{rgb}{0.13,0.13,0.13}
\definecolor{grey14}{rgb}{0.14,0.14,0.14}
\definecolor{grey15}{rgb}{0.15,0.15,0.15}
\definecolor{grey16}{rgb}{0.16,0.16,0.16}
\definecolor{grey17}{rgb}{0.17,0.17,0.17}
\definecolor{grey18}{rgb}{0.18,0.18,0.18}
\definecolor{grey19}{rgb}{0.19,0.19,0.19}
\definecolor{grey1}{rgb}{0.01,0.01,0.01}
\definecolor{grey20}{rgb}{0.20,0.20,0.20}
\definecolor{grey21}{rgb}{0.21,0.21,0.21}
\definecolor{grey22}{rgb}{0.22,0.22,0.22}
\definecolor{grey23}{rgb}{0.23,0.23,0.23}
\definecolor{grey24}{rgb}{0.24,0.24,0.24}
\definecolor{grey25}{rgb}{0.25,0.25,0.25}
\definecolor{grey26}{rgb}{0.26,0.26,0.26}
\definecolor{grey27}{rgb}{0.27,0.27,0.27}
\definecolor{grey28}{rgb}{0.28,0.28,0.28}
\definecolor{grey29}{rgb}{0.29,0.29,0.29}
\definecolor{grey2}{rgb}{0.02,0.02,0.02}
\definecolor{grey30}{rgb}{0.30,0.30,0.30}
\definecolor{grey31}{rgb}{0.31,0.31,0.31}
\definecolor{grey32}{rgb}{0.32,0.32,0.32}
\definecolor{grey33}{rgb}{0.33,0.33,0.33}
\definecolor{grey34}{rgb}{0.34,0.34,0.34}
\definecolor{grey35}{rgb}{0.35,0.35,0.35}
\definecolor{grey36}{rgb}{0.36,0.36,0.36}
\definecolor{grey37}{rgb}{0.37,0.37,0.37}
\definecolor{grey38}{rgb}{0.38,0.38,0.38}
\definecolor{grey39}{rgb}{0.39,0.39,0.39}
\definecolor{grey3}{rgb}{0.03,0.03,0.03}
\definecolor{grey40}{rgb}{0.40,0.40,0.40}
\definecolor{grey41}{rgb}{0.41,0.41,0.41}
\definecolor{grey42}{rgb}{0.42,0.42,0.42}
\definecolor{grey43}{rgb}{0.43,0.43,0.43}
\definecolor{grey44}{rgb}{0.44,0.44,0.44}
\definecolor{grey45}{rgb}{0.45,0.45,0.45}
\definecolor{grey46}{rgb}{0.46,0.46,0.46}
\definecolor{grey47}{rgb}{0.47,0.47,0.47}
\definecolor{grey48}{rgb}{0.48,0.48,0.48}
\definecolor{grey49}{rgb}{0.49,0.49,0.49}
\definecolor{grey4}{rgb}{0.04,0.04,0.04}
\definecolor{grey50}{rgb}{0.50,0.50,0.50}
\definecolor{grey51}{rgb}{0.51,0.51,0.51}
\definecolor{grey52}{rgb}{0.52,0.52,0.52}
\definecolor{grey53}{rgb}{0.53,0.53,0.53}
\definecolor{grey54}{rgb}{0.54,0.54,0.54}
\definecolor{grey55}{rgb}{0.55,0.55,0.55}
\definecolor{grey56}{rgb}{0.56,0.56,0.56}
\definecolor{grey57}{rgb}{0.57,0.57,0.57}
\definecolor{grey58}{rgb}{0.58,0.58,0.58}
\definecolor{grey59}{rgb}{0.59,0.59,0.59}
\definecolor{grey5}{rgb}{0.05,0.05,0.05}
\definecolor{grey60}{rgb}{0.60,0.60,0.60}
\definecolor{grey61}{rgb}{0.61,0.61,0.61}
\definecolor{grey62}{rgb}{0.62,0.62,0.62}
\definecolor{grey63}{rgb}{0.63,0.63,0.63}
\definecolor{grey64}{rgb}{0.64,0.64,0.64}
\definecolor{grey65}{rgb}{0.65,0.65,0.65}
\definecolor{grey66}{rgb}{0.66,0.66,0.66}
\definecolor{grey67}{rgb}{0.67,0.67,0.67}
\definecolor{grey68}{rgb}{0.68,0.68,0.68}
\definecolor{grey69}{rgb}{0.69,0.69,0.69}
\definecolor{grey6}{rgb}{0.06,0.06,0.06}
\definecolor{grey70}{rgb}{0.70,0.70,0.70}
\definecolor{grey71}{rgb}{0.71,0.71,0.71}
\definecolor{grey72}{rgb}{0.72,0.72,0.72}
\definecolor{grey73}{rgb}{0.73,0.73,0.73}
\definecolor{grey74}{rgb}{0.74,0.74,0.74}
\definecolor{grey75}{rgb}{0.75,0.75,0.75}
\definecolor{grey76}{rgb}{0.76,0.76,0.76}
\definecolor{grey77}{rgb}{0.77,0.77,0.77}
\definecolor{grey78}{rgb}{0.78,0.78,0.78}
\definecolor{grey79}{rgb}{0.79,0.79,0.79}
\definecolor{grey7}{rgb}{0.07,0.07,0.07}
\definecolor{grey80}{rgb}{0.80,0.80,0.80}
\definecolor{grey81}{rgb}{0.81,0.81,0.81}
\definecolor{grey82}{rgb}{0.82,0.82,0.82}
\definecolor{grey83}{rgb}{0.83,0.83,0.83}
\definecolor{grey84}{rgb}{0.84,0.84,0.84}
\definecolor{grey85}{rgb}{0.85,0.85,0.85}
\definecolor{grey86}{rgb}{0.86,0.86,0.86}
\definecolor{grey87}{rgb}{0.87,0.87,0.87}
\definecolor{grey88}{rgb}{0.88,0.88,0.88}
\definecolor{grey89}{rgb}{0.89,0.89,0.89}
\definecolor{grey8}{rgb}{0.08,0.08,0.08}
\definecolor{grey90}{rgb}{0.90,0.90,0.90}
\definecolor{grey91}{rgb}{0.91,0.91,0.91}
\definecolor{grey92}{rgb}{0.92,0.92,0.92}
\definecolor{grey93}{rgb}{0.93,0.93,0.93}
\definecolor{grey94}{rgb}{0.94,0.94,0.94}
\definecolor{grey95}{rgb}{0.95,0.95,0.95}
\definecolor{grey96}{rgb}{0.96,0.96,0.96}
\definecolor{grey97}{rgb}{0.97,0.97,0.97}
\definecolor{grey98}{rgb}{0.98,0.98,0.98}
\definecolor{grey99}{rgb}{0.99,0.99,0.99}
\definecolor{grey9}{rgb}{0.09,0.09,0.09}
\definecolor{grey}{rgb}{0.75,0.75,0.75}
\definecolor{honeydew1}{rgb}{0.94,1.00,0.94}
\definecolor{honeydew2}{rgb}{0.88,0.93,0.88}
\definecolor{honeydew3}{rgb}{0.76,0.80,0.76}
\definecolor{honeydew4}{rgb}{0.51,0.55,0.51}
\definecolor{honeydew}{rgb}{0.94,1.00,0.94}
\definecolor{hotpink}{rgb}{1.00,0.41,0.71}
\definecolor{indianred}{rgb}{0.80,0.36,0.36}
\definecolor{ivory1}{rgb}{1.00,1.00,0.94}
\definecolor{ivory2}{rgb}{0.93,0.93,0.88}
\definecolor{ivory3}{rgb}{0.80,0.80,0.76}
\definecolor{ivory4}{rgb}{0.55,0.55,0.51}
\definecolor{ivory}{rgb}{1.00,1.00,0.94}
\definecolor{khaki1}{rgb}{1.00,0.96,0.56}
\definecolor{khaki2}{rgb}{0.93,0.90,0.52}
\definecolor{khaki3}{rgb}{0.80,0.78,0.45}
\definecolor{khaki4}{rgb}{0.55,0.53,0.31}
\definecolor{khaki}{rgb}{0.94,0.90,0.55}
\definecolor{lavenderblush}{rgb}{1.00,0.94,0.96}
\definecolor{lavender}{rgb}{0.90,0.90,0.98}
\definecolor{lawngreen}{rgb}{0.49,0.99,0.00}
\definecolor{lemonchiffon}{rgb}{1.00,0.98,0.80}
\definecolor{lightblue}{rgb}{0.68,0.85,0.90}
\definecolor{lightcoral}{rgb}{0.94,0.50,0.50}
\definecolor{lightcyan}{rgb}{0.88,1.00,1.00}
\definecolor{lightgoldenrod}{rgb}{0.93,0.87,0.51}
\definecolor{lightgoldenrod}{rgb}{0.98,0.98,0.82}
\definecolor{lightgray}{rgb}{0.83,0.83,0.83}
\definecolor{lightgreen}{rgb}{0.56,0.93,0.56}
\definecolor{lightgrey}{rgb}{0.83,0.83,0.83}
\definecolor{lightpink}{rgb}{1.00,0.71,0.76}
\definecolor{lightsalmon}{rgb}{1.00,0.63,0.48}
\definecolor{lightsea}{rgb}{0.13,0.70,0.67}
\definecolor{lightsky}{rgb}{0.53,0.81,0.98}
\definecolor{lightslate}{rgb}{0.47,0.53,0.60}
\definecolor{lightslate}{rgb}{0.47,0.53,0.60}
\definecolor{lightslate}{rgb}{0.52,0.44,1.00}
\definecolor{lightsteel}{rgb}{0.69,0.77,0.87}
\definecolor{lightyellow}{rgb}{1.00,1.00,0.88}
\definecolor{limegreen}{rgb}{0.20,0.80,0.20}
\definecolor{linen}{rgb}{0.98,0.94,0.90}
\definecolor{magenta1}{rgb}{1.00,0.00,1.00}
\definecolor{magenta2}{rgb}{0.93,0.00,0.93}
\definecolor{magenta3}{rgb}{0.80,0.00,0.80}
\definecolor{magenta4}{rgb}{0.55,0.00,0.55}
\definecolor{magenta}{rgb}{1.00,0.00,1.00}
\definecolor{maroon1}{rgb}{1.00,0.20,0.70}
\definecolor{maroon2}{rgb}{0.93,0.19,0.65}
\definecolor{maroon3}{rgb}{0.80,0.16,0.56}
\definecolor{maroon4}{rgb}{0.55,0.11,0.38}
\definecolor{maroon}{rgb}{0.69,0.19,0.38}
\definecolor{mediumaquamarine}{rgb}{0.40,0.80,0.67}
\definecolor{mediumblue}{rgb}{0.00,0.00,0.80}
\definecolor{mediumorchid}{rgb}{0.73,0.33,0.83}
\definecolor{mediumpurple}{rgb}{0.58,0.44,0.86}
\definecolor{mediumsea}{rgb}{0.24,0.70,0.44}
\definecolor{mediumslate}{rgb}{0.48,0.41,0.93}
\definecolor{mediumspring}{rgb}{0.00,0.98,0.60}
\definecolor{mediumturquoise}{rgb}{0.28,0.82,0.80}
\definecolor{mediumviolet}{rgb}{0.78,0.08,0.52}
\definecolor{midnightblue}{rgb}{0.10,0.10,0.44}
\definecolor{mintcream}{rgb}{0.96,1.00,0.98}
\definecolor{mistyrose}{rgb}{1.00,0.89,0.88}
\definecolor{moccasin}{rgb}{1.00,0.89,0.71}
\definecolor{navajowhite}{rgb}{1.00,0.87,0.68}
\definecolor{navyblue}{rgb}{0.00,0.00,0.50}
\definecolor{navy}{rgb}{0.00,0.00,0.50}
\definecolor{oldlace}{rgb}{0.99,0.96,0.90}
\definecolor{olivedrab}{rgb}{0.42,0.56,0.14}
\definecolor{orange1}{rgb}{1.00,0.65,0.00}
\definecolor{orange2}{rgb}{0.93,0.60,0.00}
\definecolor{orange3}{rgb}{0.80,0.52,0.00}
\definecolor{orange4}{rgb}{0.55,0.35,0.00}
\definecolor{orangered}{rgb}{1.00,0.27,0.00}
\definecolor{orange}{rgb}{1.00,0.65,0.00}
\definecolor{orchid1}{rgb}{1.00,0.51,0.98}
\definecolor{orchid2}{rgb}{0.93,0.48,0.91}
\definecolor{orchid3}{rgb}{0.80,0.41,0.79}
\definecolor{orchid4}{rgb}{0.55,0.28,0.54}
\definecolor{orchid}{rgb}{0.85,0.44,0.84}
\definecolor{palegoldenrod}{rgb}{0.93,0.91,0.67}
\definecolor{palegreen}{rgb}{0.60,0.98,0.60}
\definecolor{paleturquoise}{rgb}{0.69,0.93,0.93}
\definecolor{paleviolet}{rgb}{0.86,0.44,0.58}
\definecolor{papayawhip}{rgb}{1.00,0.94,0.84}
\definecolor{peachpuff}{rgb}{1.00,0.85,0.73}
\definecolor{peru}{rgb}{0.80,0.52,0.25}
\definecolor{pink1}{rgb}{1.00,0.71,0.77}
\definecolor{pink2}{rgb}{0.93,0.66,0.72}
\definecolor{pink3}{rgb}{0.80,0.57,0.62}
\definecolor{pink4}{rgb}{0.55,0.39,0.42}
\definecolor{pink}{rgb}{1.00,0.75,0.80}
\definecolor{plum1}{rgb}{1.00,0.73,1.00}
\definecolor{plum2}{rgb}{0.93,0.68,0.93}
\definecolor{plum3}{rgb}{0.80,0.59,0.80}
\definecolor{plum4}{rgb}{0.55,0.40,0.55}
\definecolor{plum}{rgb}{0.87,0.63,0.87}
\definecolor{powderblue}{rgb}{0.69,0.88,0.90}
\definecolor{purple1}{rgb}{0.61,0.19,1.00}
\definecolor{purple2}{rgb}{0.57,0.17,0.93}
\definecolor{purple3}{rgb}{0.49,0.15,0.80}
\definecolor{purple4}{rgb}{0.33,0.10,0.55}
\definecolor{purple}{rgb}{0.63,0.13,0.94}
\definecolor{red1}{rgb}{1.00,0.00,0.00}
\definecolor{red2}{rgb}{0.93,0.00,0.00}
\definecolor{red3}{rgb}{0.80,0.00,0.00}
\definecolor{red4}{rgb}{0.55,0.00,0.00}
\definecolor{red}{rgb}{1.00,0.00,0.00}
\definecolor{rosybrown}{rgb}{0.74,0.56,0.56}
\definecolor{royalblue}{rgb}{0.25,0.41,0.88}
\definecolor{saddlebrown}{rgb}{0.55,0.27,0.07}
\definecolor{salmon1}{rgb}{1.00,0.55,0.41}
\definecolor{salmon2}{rgb}{0.93,0.51,0.38}
\definecolor{salmon3}{rgb}{0.80,0.44,0.33}
\definecolor{salmon4}{rgb}{0.55,0.30,0.22}
\definecolor{salmon}{rgb}{0.98,0.50,0.45}
\definecolor{sandybrown}{rgb}{0.96,0.64,0.38}
\definecolor{seagreen}{rgb}{0.18,0.55,0.34}
\definecolor{seashell1}{rgb}{1.00,0.96,0.93}
\definecolor{seashell2}{rgb}{0.93,0.90,0.87}
\definecolor{seashell3}{rgb}{0.80,0.77,0.75}
\definecolor{seashell4}{rgb}{0.55,0.53,0.51}
\definecolor{seashell}{rgb}{1.00,0.96,0.93}
\definecolor{sienna1}{rgb}{1.00,0.51,0.28}
\definecolor{sienna2}{rgb}{0.93,0.47,0.26}
\definecolor{sienna3}{rgb}{0.80,0.41,0.22}
\definecolor{sienna4}{rgb}{0.55,0.28,0.15}
\definecolor{sienna}{rgb}{0.63,0.32,0.18}
\definecolor{skyblue}{rgb}{0.53,0.81,0.92}
\definecolor{slateblue}{rgb}{0.42,0.35,0.80}
\definecolor{slategray}{rgb}{0.44,0.50,0.56}
\definecolor{slategrey}{rgb}{0.44,0.50,0.56}
\definecolor{snow1}{rgb}{1.00,0.98,0.98}
\definecolor{snow2}{rgb}{0.93,0.91,0.91}
\definecolor{snow3}{rgb}{0.80,0.79,0.79}
\definecolor{snow4}{rgb}{0.55,0.54,0.54}
\definecolor{snow}{rgb}{1.00,0.98,0.98}
\definecolor{springgreen}{rgb}{0.00,1.00,0.50}
\definecolor{steelblue}{rgb}{0.27,0.51,0.71}
\definecolor{tan1}{rgb}{1.00,0.65,0.31}
\definecolor{tan2}{rgb}{0.93,0.60,0.29}
\definecolor{tan3}{rgb}{0.80,0.52,0.25}
\definecolor{tan4}{rgb}{0.55,0.35,0.17}
\definecolor{tan}{rgb}{0.82,0.71,0.55}
\definecolor{thistle1}{rgb}{1.00,0.88,1.00}
\definecolor{thistle2}{rgb}{0.93,0.82,0.93}
\definecolor{thistle3}{rgb}{0.80,0.71,0.80}
\definecolor{thistle4}{rgb}{0.55,0.48,0.55}
\definecolor{thistle}{rgb}{0.85,0.75,0.85}
\definecolor{tomato1}{rgb}{1.00,0.39,0.28}
\definecolor{tomato2}{rgb}{0.93,0.36,0.26}
\definecolor{tomato3}{rgb}{0.80,0.31,0.22}
\definecolor{tomato4}{rgb}{0.55,0.21,0.15}
\definecolor{tomato}{rgb}{1.00,0.39,0.28}
\definecolor{turquoise1}{rgb}{0.00,0.96,1.00}
\definecolor{turquoise2}{rgb}{0.00,0.90,0.93}
\definecolor{turquoise3}{rgb}{0.00,0.77,0.80}
\definecolor{turquoise4}{rgb}{0.00,0.53,0.55}
\definecolor{turquoise}{rgb}{0.25,0.88,0.82}
\definecolor{violetred}{rgb}{0.82,0.13,0.56}
\definecolor{violet}{rgb}{0.93,0.51,0.93}
\definecolor{wheat1}{rgb}{1.00,0.91,0.73}
\definecolor{wheat2}{rgb}{0.93,0.85,0.68}
\definecolor{wheat3}{rgb}{0.80,0.73,0.59}
\definecolor{wheat4}{rgb}{0.55,0.49,0.40}
\definecolor{wheat}{rgb}{0.96,0.87,0.70}
\definecolor{whitesmoke}{rgb}{0.96,0.96,0.96}
\definecolor{white}{rgb}{1.00,1.00,1.00}
\definecolor{yellow1}{rgb}{1.00,1.00,0.00}
\definecolor{yellow2}{rgb}{0.93,0.93,0.00}
\definecolor{yellow3}{rgb}{0.80,0.80,0.00}
\definecolor{yellow4}{rgb}{0.55,0.55,0.00}
\definecolor{yellowgreen}{rgb}{0.60,0.80,0.20}
\definecolor{yellow}{rgb}{1.00,1.00,0.00}

\usepackage{ulem}       

\begin{document}        


\maketitle
\label{firstpage}

\begin{abstract}
{\color{black}

We present a catalogue of 39 multiple-mergers found using the mergers catalogue of the Galaxy Zoo project for $z<0.1$ and compare them to corresponding semi-analytical galaxies from the Millennium Simulation. We estimate the (volume-limited) multi-merger fraction of the local Universe using our sample and find it to be at least two orders of magnitude less than binary-mergers - in good agreement with the simulations (especially the Munich group). We then investigate the properties of galaxies in binary- and multi-mergers (morphologies, colours, stellar masses and environment) and compare these results with those predicted by the semi-analytical galaxies. We find that multi-mergers favour galaxies with properties typical of elliptical morphologies and that this is in qualitative agreement with the models. Studies of multi-mergers thus provide an independent (and largely corroborating) test of the Millennium semi-analytical models.

}
\end{abstract}



\section{Introduction}

\begin{center}
\begin{figure*}
	\includegraphics[width=170mm]{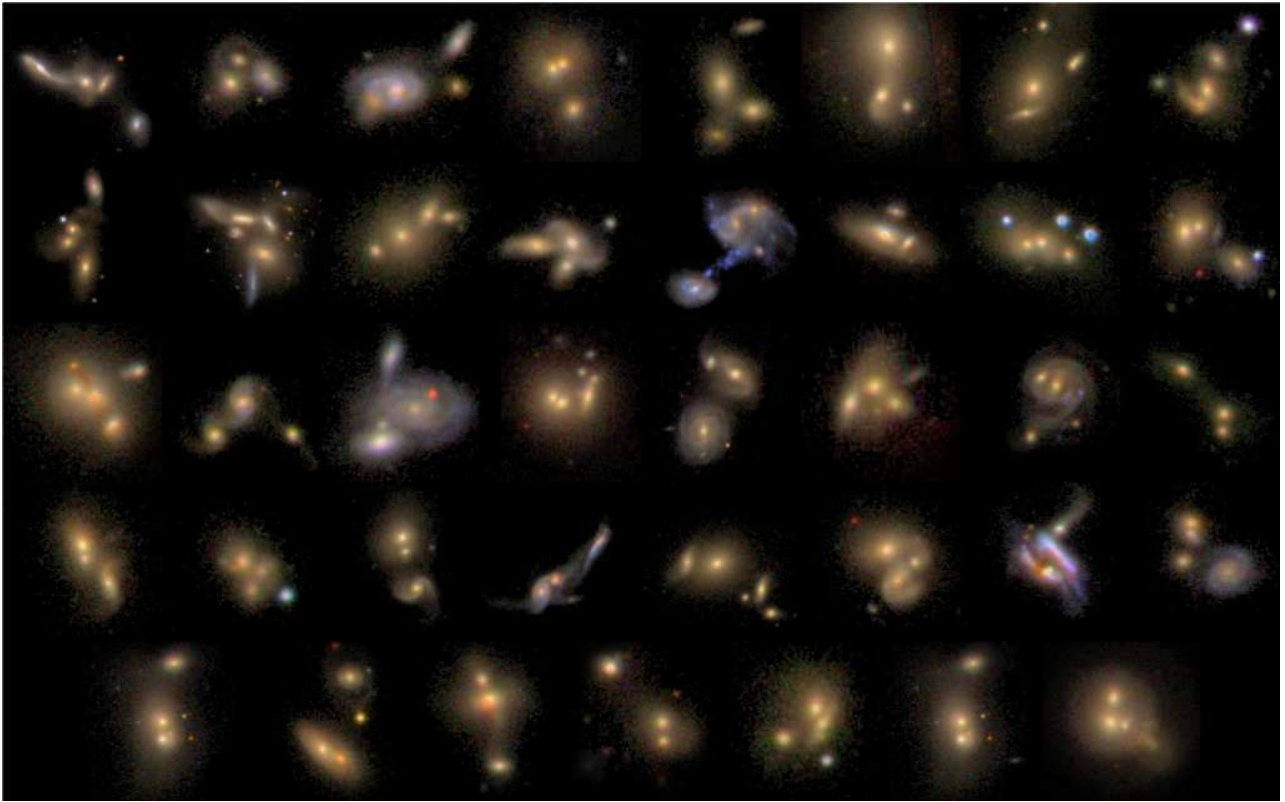}
	\caption[Figure 1]{Images of the 39 multi-merger systems obtained from GZM1. Each tile has been scaled for optimal viewing, typically with sides of $\sim 50 ''$.}
	\label{collation}
\end{figure*}
\end{center}

{\color{black}The exact extent to which mergers are able to account for the observed properties and morphologies of galaxies in the Universe preoccupies much of modern research (see \citealt{darg1} hereafter D10a and \citealt{darg2} hereafter D10b and references therein). If mergers are not sufficient to explain all observations pertaining to mass-assembly, then other modes of galaxy formation must be conjoined to the standard hierarchical scheme. If, however, mergers alone (implemented via a model of structure formation such as $\Lambda \mbox{CDM}$) are enough to explain what we observe in galaxies, then such additional processes can (and therefore should) be discarded. In either case, mergers are known to occur at significant rates (\citealt{con2008}; \citealt{lotz2}; \citealt{patton2008}; \citealt{bertone2}; \citealt{stewart1}; \citealt{robaina}) and are an indispensable explanatory resource with respect to galaxy evolution.

To this end the Galaxy Zoo project has established the largest homogeneous catalogue of merging galaxies in the local Universe with 3003 visually confirmed mergers (D10a). In this paper we examine an interesting subset of this mergers catalogue - 39 systems with three or more galaxies merging simultaneously (shown in Figure \ref{collation}) - and use it to estimate the (major) multi-merger fraction of galaxies in the local Universe. This fraction and the examined properties of these multi-merging galaxies thus provide a useful {\it independent} test of semi-analytic models (which are currently indispensable aids to the study of galaxy evolution) from the typical observables that such models are tuned to reproduce.}

In the $\Lambda \mbox{CDM}$ cosmology galactic mergers are primarily driven by the coalescence of Dark Matter (DM) structures {\color{black}(see e.g. \citealt{white2}; \citealt{bond};  \citealt{white}; \citealt{lacey}; \citealt{jenkins}; \citealt{springel}; \citealt{stewart2}).} Once DM haloes have become virialized, further growth can only occur through accretion and merging (\citealt{fakhouri}; \citealt{neistein}; \citealt{fakhouri2}). The merger histories of DM halos can be reproduced through $N-$Body simulations (\citealt{springel}; \citealt{deLucia1b}; \citealt{harker})\footnote{While the $N$-Body simulations are more or less `exact' the scheme for grouping DM-particles into halos and sub-halos is arbitrary to a certain extent, in other words, the merger history depends on the detailed definition of a halo.} or through analytic approximations such as Press-Schechter (\citealt{press}) and its extensions (\citealt{bond}; \citealt{bower1}). 

To empirically test the accuracy of DM structure formation, semi-analytic recipes for the evolution of the visible matter within these haloes is required. These aim to capture the macro-physical processes affecting observable quantities such as the Luminosity Function (\citealt{benson2}). Each physical process typically involves one or two free parameters and so semi-analytic models (SAMs) require fine-tuning to reproduce empirical observations. The major considerations are photoionisation (\citealt{benson1}), shock heating of gas (\citealt{cattaneo}), gas cooling (\citealt{deLucia4}), AGN feedback (\citealt{bower2}; \citealt{croton}), supernovae feedback (also enriching the IGM with metal; \citealt{deLucia1a}) and mergers (\citealt{springel2001}). The resultant `galaxies' that occupy the DM haloes can then be converted to observable luminosities through stellar-population synthesis models (e.g. \citealt{bruzual}; \citealt{maraston1}; \citealt{maraston2}).  

Given enough adjustment of the free-parameter values and phenomenological ingredients to represent time-dependant feedback effects, any one quantity such as the empirically determined Luminosity Function (\citealt{cole2001}; \citealt{norberg}; \citealt{huang}; \citealt{panter}; \citealt{jones}; \citealt{devereux}), can be approximated arbitrarily well {\color{black} in principle}. A major test of a model's veracity therefore rests in its capacity to reproduce observables that were {\it not} involved in it's original calibration. One such test is to determine how well a SAM agrees with observed merger rates (or fractions). Several studies have compared the evolution of the galaxy merger rate obtained by the close-pairs technique (see \S \ref{method} for description) to that of SAMs implemented in the Millennium Run (\citealt{kitz}; \citealt{patton2008}; \citealt{mateus}; \citealt{hopkins3}). \citealt{bertone2} more recently compared the SAM of \citealt{bertone1} to the merger rate obtained by the CAS {\color{black}(concentration, asymmetry and clumpiness;} \citealt{con2003a}; \citealt{con2003b}) method and found the two were roughly consistent for $z \lesssim 2$. 

In this paper, we introduce a new test for galaxy-evolution models that has, until now, been too difficult to find observationally: the fraction and properties of galaxies in (near simultaneous) multi-mergers (mergers of three or more galaxies of similar mass). Several individual multi-merger systems have been studied (\citealt{cui}; \citealt{amram}; \citealt{rines}) and a few numerical simulations of multi-mergers carried out (\citealt{weil}; \citealt{bekki}; {\color{black}\citealt{renaud}}) but no practical method has been obtained till now that might locate multi-mergers in a near complete manner.

In \S \ref{catalogue_construction} we describe how the Galaxy Zoo project was able to construct such a catalogue in order to estimate the multi-merger fraction (carried out in \S \ref{mm_fraction_sdss}). We then compare the binary and multi-merger fractions obtained by the Galaxy Zoo project with those of the SAMs of the Millennium. They fall broadly into two families of models: those developed by MPA Garching (\citealt{croton}; \citealt{deLucia2}; \citealt{bertone1}) and those of Durham (\citealt{cole}; \citealt{benson1}; \citealt{benson2}; \citealt{baugh}; \citealt{bower2}; {\color{black}\citealt{font}}). Several implementations have been developed by both groups and we use those which are publicly available.\footnote{See http://www.mpa-garching.mpg.de/millennium/} For the MPA, these are the models of \citealt{deLucia2} \& \citealt{deLucia3} (hereafter delucia06) and \citealt{bertone1} (hereafter bertone07, an extension of delucia06) and for Durham, the model of \citealt{bower2} (hereafter bower06). {\color{black} We describe some of their characteristics in \S \ref{mill}.}

By comparing the SAM merger fractions to SDSS observations made herein (\S \ref{mm_fraction}), we effectively test the accuracy of the build up of clumpiness in the Universe since the main factor affecting merger rates is environment. Other properties correlate with environment such as stellar mass, colour and morphology and so we examine these in both the SAMs and Galaxy Zoo catalogues for multi-mergers, binary-mergers and single galaxy systems (\S \ref{properties}). We summarise our results in \S \ref{summary}.

\begin{center}
\begin{figure}
	\includegraphics[width=84mm]{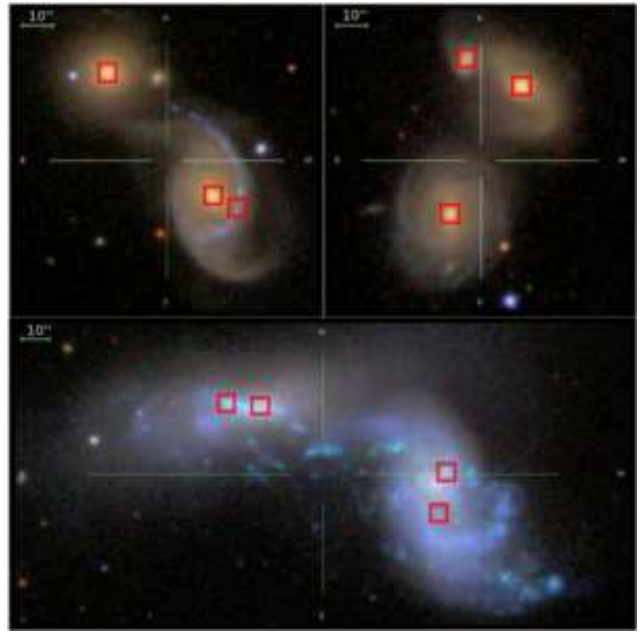}
	\caption[Figure 1]{Examples of possible distributions of SDSS objects with spectra (red squares) in merging systems. The close-pairs technique in particular would be unable to distinguish which of these systems is a `multi-merger' (for example, the top two panels have three spectral objects each but the the left-hand system is a binary-merger while the right-hand system is a triple-merger; the system on the lower panel is a binary-merger but has {\it four} spectral objects).}
	\label{examples}
\end{figure}
\end{center}


\begin{table*}
\begin{minipage}{126mm}
	\caption[Table 1]{Catalogue of multi-mergers. `SDSS Objid' is the unique label for the primary (object chosen by Galaxy Zoo) for each system. `Type' decribes whether the multi-merger is major, middle or minor (see text for definition). $\mbox{M}_1$, $\mbox{M}_2$, etc. give the stellar mass estimates of the galaxies in descending order for the system.	Galaxies marked * are part of a {\it major} triple-merger ($\mbox{M}_1/\mbox{M}_2<3$ and $\mbox{M}_2/\mbox{M}_3<3$) {\it and} are bright enough to be in the volume-limited analysis ($M_r<-20.55$) thereby contributing to the major triple-merger fraction (see \S \ref{mm_fraction}). The entry marked ! has no photometric readings and shows a rare failure of the SDSS deblending routine.}	
	\label{table1}
	\begin{tabular}{| c | c | c | c | c | c | c | c | c | c |}
\hline \\
 & Type & SDSS Objid & M1 & M2 & M3 & M4 & M5 & M1 \slash M2 & M2 \slash M3 \\
  &  &  & \multicolumn{5}{ c }{ $ (x \in 10^{x} \mbox{M}_{\astrosun}) $}  &  &  \\
\hline \\

       1 & Minor & 587725491062571042 & 11.6 & 10.7 & 9.94 &  &  & 8.06 & 5.94 \\ 
       2 & Major & 587725816951865440 & 11.6* & 11.5* & 11.0* &  &  & 1.20 & 2.65 \\ 
       3 & Major & 587726014546772187 & 11.7* & 11.6* & 11.5* & 11.0 &  & 1.26 & 1.51 \\ 
       4 & Minor & 587726016157384716 & 11.0 & 10.4 & 10.3 &  &  & 4.21 & 1.10 \\ 
       5 & Major & 587726033303765102 & 11.7* & 11.4* & 10.9* &  &  & 1.98 & 2.88 \\ 
       6 & Middle & 587726033846009923 & 11.7 & 11.4 & 10.5 &  &  & 2.17 & 7.38 \\ 
       7 & Minor & 587727865644974418 & 11.3 & 10.7 & 10.1 &  &  & 3.51 & 3.99 \\ 
       8 & Minor & 587729385547038748 & 12.0 & 11.1 & 10.9 & 10.2 &  & 7.56 & 1.58 \\ 
       9 & Minor & 587729772070633570 & 11.6 & 10.6 & 10.5 &  &  & 8.96 & 1.44 \\ 
      10 & Minor & 587731511544905794 & 11.9 & 10.9 & ! &  &  & 8.42 & NA \\ 
      11 & Major & 587732050018107546 & 11.1* & 10.7 & 10.6* &  &  & 2.66 & 1.08 \\ 
      12 & Major & 587732483809345720 & 11.5* & 11.4* & 11.2* &  &  & 1.31 & 1.84 \\ 
      13 & Minor & 587732483811770549 & 11.7 & 10.7 & 10.6 & 9.99 &  & 10.4 & 1.23 \\ 
      14 & Minor & 587733081346605098 & 10.9 & 10.0 & 9.18 &  &  & 8.47 & 6.71 \\ 
      15 & Major & 587734303805604056 & 11.7* & 11.4* & 11.3* &  &  & 2.01 & 1.36 \\ 
      16 & Minor & 587735696979656747 & 10.7 & 10.2 & 9.62 &  &  & 3.18 & 4.58 \\ 
      17 & Minor & 587736477053681696 & 11.5 & 10.7 & 10.4 &  &  & 6.79 & 2.00 \\ 
      18 & Major & 587736807771078935 & 11.7* & 11.3* & 11.2* &  &  & 2.28 & 1.26 \\ 
      19 & Major & 587736919972643151 & 11.9* & 11.7* & 11.2* & 10.6* &  & 1.93 & 2.60 \\ 
      20 & Middle & 587738952027734025 & 11.6 & 11.2 & 10.1 &  &  & 2.48 & 11.9 \\ 
      21 & Major & 587739096450727993 & 9.98 & 9.83 & 9.48 &  &  & 1.40 & 2.24 \\ 
      22 & Middle & 587739382058909818 & 11.5 & 11.2 & 10.5 &  &  & 2.14 & 4.57 \\ 
      23 & Middle & 587739607547904036 & 11.2 & 11.2 & 9.90 &  &  & 1.12 & 20.2 \\ 
      24 & Minor & 587739646203461677 & 11.8 & 10.8 & 10.0 &  &  & 9.95 & 6.71 \\ 
      25 & Major & 587739652645191695 & 10.4 & 10.2 & 10.2 & 9.38 & 8.58 & 1.62 & 1.02 \\ 
      26 & Middle & 587739811038101752 & 11.4 & 11.1 & 10.3 &  &  & 2.33 & 6.03 \\ 
      27 & Minor & 587739844321542224 & 11.4 & 10.8 & 10.8 & 10.6 &  & 3.62 & 1.03 \\ 
      28 & Major & 587742014603985009 & 11.3* & 11.1* & 11.1* &  &  & 1.44 & 1.08 \\ 
      29 & Middle & 587742060517064897 & 11.9 & 11.4 & 10.9 &  &  & 2.59 & 3.16 \\ 
      30 & Major & 587742062126235884 & 11.1* & 10.9 & 10.4 &  &  & 1.80 & 2.75 \\ 
      31 & Minor & 587744727686381602 & 11.3 & 10.8 & 10.2 &  &  & 3.54 & 3.68 \\ 
      32 & Middle & 587745244697329759 & 11.5 & 11.4 & 10.4 &  &  & 1.18 & 11.0 \\ 
      33 & Major & 588010359624826929 & 11.6* & 11.3* & 11.0* &  &  & 2.15 & 2.01 \\ 
      34 & Minor & 588013382728221044 & 11.5 & 10.9 & 10.6 &  &  & 4.31 & 2.18 \\ 
      35 & Middle & 588013384351678489 & 11.1 & 11.0 & 10.3 & 10.0 &  & 1.14 & 5.60 \\ 
      36 & Major & 588016878292631762 & 11.4* & 11.0* & 10.6 &  &  & 2.39 & 2.65 \\ 
      37 & Minor & 588017705070100599 & 11.4 & 10.7 & 10.6 &  &  & 5.15 & 1.19 \\ 
      38 & Major & 588018090007003216 & 11.6* & 11.3* & 11.1 &  &  & 1.90 & 1.58 \\ 
      39 & Major & 588848898849439845 & 11.0* & 10.7* & 10.2* &  &  & 1.74 & 2.98 \\ 

\end{tabular}
\end{minipage}
\end{table*}

\section{The Multi-Merger Catalogue}
\label{catalogue_construction}
\subsection{Finding Multi-Mergers}
\label{method}

As discussed in D10a, finding mergers amongst surveys with $\sim 10^6$ galaxies is highly non-trivial. Non-parametric techniques such as CAS and GM$_{20}$ (Gini coefficient and the second-order moment of the brightest $20\%$ of the galaxy's light; \citealt{abraham}; \citealt{lotz1}; \citealt{lotz2}) that aim to identify parameter spaces uniquely occupied by mergers have thus far proved challenging and the prospect of finding even more specific sub-spaces limited solely to {\it multi}-mergers is unrealistic (\citealt{lisker}; D10a).

Likewise, modifying the close-pairs technique {\color{black}(locating galaxy pairs within a certain angular separation and redshift difference)} to find multi-mergers within SDSS is impractical due to fibre overlaps. The apparatus within SDSS will not acquire spectra (which are needed in such `blind' methods to avoid projection effects) from objects within $55''$ of each other on a single viewing. The conventional close-pairs technique is therefore only useful within tile-overlap regions (\citealt{strauss}; \citealt{blanton2}; D10a). In order to find systems with {\it three} or more galaxies in the merger stage which are within $55''$ of each other, the system would have to rest in part of the sky where there is a {\it double} tile-overlap. 

We investigated the practicality of using a modified close-pairs technique to find multi-mergers using a close-pairs catalogue restricted to $0.005<z<0.1$ that associates spectral galaxy objects with a redshift tolerance of $0.0017$ {\color{black}(corresponding to a velocity difference of $500$kms$^{-1}$ as used in \citealt{patton2002})} and projected separation of 30 kpc. We found that of the $4880$ spectral objects in this catalogue, only 48 systems (148 objects) had three or more spectral objects. Only one of these 48 systems belonged to our catalogue {\color{black}indicating how incomplete this technique is for finding multi-mergers. Even when three spectral objects do fall within these limits, the close-pairs technique would not distinguish multi-mergers from normal binary-mergers that have been deblended with multiple spectral objects. Figure \ref{examples} shows examples of this problem - the upper images both have three spectral objects, but the left image is a binary-merger (with two spectral objects on one galaxy) and the right image is a triple-merger. Likewise, without visual inspection, it would be almost impossible to know if the lower image, which has four spectral objects, contained two, three or four galaxies interacting with each other.} 

The only way to find multi-merging systems is therefore through visual inspection since humans can readily distinguish between features like bulges and tidal tails and can often, on this basis, tell straight away if a system is a multi-merger. The disadvantage with visual identification by individuals is, firstly, it is time consuming in general and wholly impractical for surveys with $\sim 10^6$ galaxies and, secondly, it is subjective. The Galaxy Zoo project helps overcome both of these problems though. It has been shown that human classifications averaged over {\it large numbers} of individuals do provide a good measure of morphological classification (\citealt{lintott1}; \citealt{lintott2}) and merger identification (D10a) and, with instruction, can often identify multi-mergers with ease.  

The only method one can use at present is therefore to visually flag any system that could be a multi-merger (three galaxies of comparable size merging simultaneously), identify the relevant photometry for each galaxy, and use this information to more objectively assess what subset of those originally flagged are genuinely of comparable size. This is explained in more detail in what follows.

\subsection{Construction of Catalogue}

The measure relevant to mergers produced by the original Galaxy Zoo interface\footnote{Note: as of February 2009, a new interface (Zoo Two) has been in operation. These results are derived from D10a whose catalogue was constructed using the data obtained in January 2008.} is called the `weighted-merger-vote fraction' ($f_m \in [0,1]$; D10a) and was used to create a catalogue of 3003 mergers called GZM1 (with the constraints $0.005<z<0.1$ and $0.4< f_m$). {\color{black}It is calculated by dividing the total number of times users classified the system as a merger multiplied by a weighting factor representing the reliability of the particular users who examined the system (a user is weighted highly if that person tends to agree with the majority opinion) and divided by the total number of classifications that the system received (see \citealt{lintott1} for details).}
	
The creation of this catalogue required the visual re-examination of all 3003 systems to check for misclassifications and to assign morphologies to the individual galaxies. During this process, any system that appeared as though it {\it might} be a multi-merger was noted for future study. 78 such systems were flagged and form the parent sample for this catalogue. Closer examination of these 78 led to the conclusion that several systems were almost certainly {\it not} merging (but were projection effects), some were too difficult to distinguish and 39 of the original 3003 systems, which make up the catalogue, are confidently multi-mergers {\color{black} with signs of interaction - most of which are discernible in Figure \ref{collation} (simple inspection of this figure should convey just how variable multi-mergers are in appearance and therefore how challenging a pattern-recognition problem it would be to design an automated system that could reliably filter these out as such).} 

Having visually determined that these 39 systems of three or more galaxies (with at least one being a Galaxy Zoo spectral object associated with an $f_m$ value $>0.4$) are multi-mergers, we manually selected the two (or more) best neighbour objects available to represent the other galaxies. The `best' object was judged according to (i) brightness in the r-band and (ii) visual common sense (similar to D10a \S 2.1). Having spectra for at least one galaxy in each multi-merging system and photometric data for each individual galaxy allows us to calculate rest-frame colours and stellar-mass approximations.

\begin{figure*}
	\includegraphics[width=150mm]{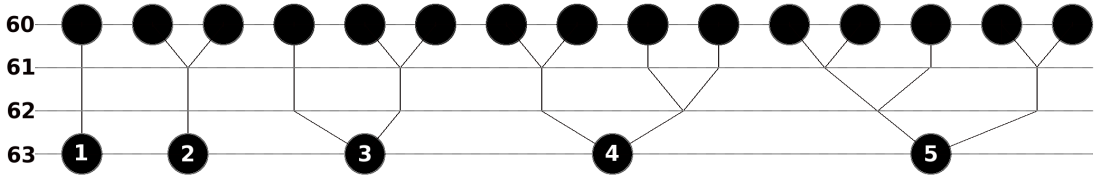}
	\caption[Figure 3]{Millennium Simulation merger-tree scheme. For all descendant galaxies at snapshot number (SN) =63 with $M_r<-20.55$, the progenitors at SN=60 are found and counted to give the number of galaxies in single-, binary- and multi-merger systems.}
	\label{tree}
\end{figure*}

\subsection{Mass and Rest-Frame Photometry Calculations}

{\color{black} To find a volume-limited major multi-merger fraction we need to calculate the rest-frame photometries and stellar masses of the galaxies in our catalogue and define what we mean by `major' in this context.}

Stellar masses are calculated as in D10a by fitting the SDSS photometries of each individual galaxy object to a library of stellar-synthesis populations (\citealt{maraston1}; \citealt{maraston2}) out to the redshift given by the spectral object in the merging system. The photometric errors given by SDSS are carried through to estimate errors for the stellar masses. K-corrected rest-frame colours are calculated as in D10b using the photometries and spectral redshifts input into the publicly available IDL routine \texttt{kcorrect 4\_1\_4} (\citealt{blanton}). The rest-frame r-band magnitude ($M_r$) is needed to volume-limit the galaxies for our analysis as in D10a and D10b.\footnote{Specifically, for an object to be capable of obtaining spectra in SDSS, it must have a petrosian magnitude of $r<17.77$ which means that an object at $z=0.1$ (the upper-boundary of our volume-limited sample) must have a rest-frame magnitude of $M_r<-20.55$.}

{\color{black} For each system we refer to the masses of each galaxy by descending order where $\mbox{M}_1$ is the most massive galaxy, $\mbox{M}_2$ the second most massive, etc. Following D10a, we call a binary-merger a major binary-merger if $\mbox{M}_2/\mbox{M}_1<3 $, else it is a minor binary-merger. Similarly, we categorise a multi-merger as a {\it major} multi-merger if each galaxy of decreasing mass is within one third of the mass of the next most massive galaxy ($\mbox{M}_2/\mbox{M}_1 < 3$, $\mbox{M}_3/\mbox{M}_2 < 3$, etc.). Of the 39 systems, only 12 are major triple-mergers by this definition. A {\it middle} multi-merger is where the two most massive galaxies are major ($\mbox{M}_2/\mbox{M}_1<3$) but the next two are minor ($\mbox{M}_3/\mbox{M}_2>3$) and a {\it minor} multi-merger is where the first two most massive galaxies form a minor binary-merger ($\mbox{M}_2/\mbox{M}_1>3$). A summary of the catalogue masses is given in Table \ref{table1}. Before estimating the major multi-merger fraction of the local Universe, we now introduce the Millennium SAMs so that we can compare the fractions derived from the simulations with the empirically observed fraction (\S \ref{mm_fraction}).}


\section{The Millennium Simulation SAMs}
\label{mill}
\subsection{Dark Matter}

The Millennium Run is an N-body Dark Matter simulation of a $\Lambda$CDM Universe. The original version uses $\sim 10^{10}$ particles in a cubic region of $500h^{-1}$Mpc sides in {\color{black} comoving} coordinates and periodic boundary conditions. The simulation is based on the cosmological parameters $\Omega_m = 0.25$, $\Omega_{\Lambda} = 0.75$, $\Omega_b = 0.045$, $h = 0.73$ and $\sigma _8 = 0.9$ (see  \citealt{springel} for details). 64 snapshots of the simulation were saved (from $z=127$ to $z=0$) and, from these, merger-trees can be constructed. However, the definition of haloes and halo substructure is a matter of convention. The MPA model defines a Friend-Of-Friend group (defined by particles linked to each other by 0.2 times the mean-particle separation) and determines substructure using the algorithm \texttt{SUBFIND} separating bound structures from unbound particles (determined by their velocity relative to their local potential).

The Durham models, following \citealt{harker}, carry out the same analysis with the addition of extra constraints on subhalo-definition designed to avoid tenuous `bridges' linking FOF groups but that will disappear in the (near) future rendering them distinct. The subhaloes of a FOF group are distinct haloes in their model if (i) the centre of the subhalo is outside twice the half-mass radius of the main halo or, (ii) the subhalo has retained more than 75 per cent of the mass since the last output time at which it was an independent halo.

The DM build-up is the most important determinant with respect to the merging of the baryonic galaxies. When a DM (satellite) halo bound to a more massive (main) halo becomes sufficiently disrupted and falls below the 20 particle limit for a structure to be defined, the countdown begins for its central galaxy to merge with that of the main halo. The exact timing depends on the SAM details relating to dynamical friction that models the decay of the satellite orbit. 

\subsection{Baryons}

It is important that the SAM accurately convert baryons to photometries if a realistic merger fraction (or rate) is to be obtained since any sensible merger fraction must be volume-limited (in redshift and magnitude) to ensure completeness. SAM recipes ultimately relate DM haloes to galaxy magnitudes in specific bands and if the prescriptions for galaxy evolution result in unrealistic stellar populations then the volume-limited fractions will be wrong.

The Durham and MPA SAMs differ on a number of details governing the formation and evolution of galaxies. Each new model attempts to capture some extra observational feature (see \citealt{parry} for a summary of the differences and the original papers for details). For example, delucia06 concentrates on the formation of brightest cluster galaxies whereas bertone07 develops the MPA model focussing on the treatment of metallicity production and exchange with the IGM. The model reports improvements in the suppression of star-formation in small haloes but at the expense of galaxy colour-bimodality. 


\subsection{Merger Trees}

The SDSS catalogue is limited to the local Universe ($z<0.1$) and, since we assume that the merger rate changes negligibly over this interval, we are only interested in local merger fractions in this paper. The Millennium Database only outputs halos and galaxies at 64 discrete time steps (referred to as `snapshots' labelled by `snapshot numbers' (SN) for 0 to 63). A schematic is shown in Figure \ref{tree} for merger trees with the progenitors at SN=60 related to descendents at SN$=63$. We approximate that a galaxy system would `look like' a merger, on average, if the progenitors of a descendent galaxy are identified as individual galaxies at a progenitor SN$_{prog}$ with redshift corresponding to a look-back time comparable to the time-scale of merger detectability. 

Of course, the vast majority of mergers will be minor-mergers whose rates are difficult to constrain observationally. Since we are interested in comparing the SAM mergers to the SDSS catalogue, we consider only progenitors with $M_r<-20.55$ and, as far as merger fractions go, we are only interested in major mergers (see \S \ref{mm_fraction}). {\color{black}In choosing an `average' or `typical' time-scale of detectability we face the difficulty that they depend on the properties of the galaxies in the merger and the merger-detection technique (\citealt{lotz3}; \citealt{lotz4}; \citealt{lotz5}; D10b). For example, \citealt{lotz3} found a median time-scale of $t_{merger} \approx 0.35\pm 0.15$Gyr for the {\it detectability} of the close-pairs technique whereas gas-rich spiral-mergers can remain detectable for much longer periods when using asymmetry techniques ($\gtrsim$ Gyr; \citealt{lotz5}). If a particular population has a high proportion of spiral galaxies compared to another, the relative merger fractions will therefore appear inflated. \citealt{bertone2} quote the merger time-scale range 0.4 Gyr - 1 Gyr based upon N-Body simulations and dynamical-friction calculations (\citealt{con2006}). We choose SN$_{prog}=60$ for the fiducial snapshot since it corresponds to the look-back time ( $z \approx 0.064 \leftrightarrow 0.6$ Gyr) closest to the median of the range of these studies ($0.35-1$ Gyr). We also study how the merger fraction varies with SN$_{prog}$ (see Table \ref{table_2}) and bear this uncertainty in mind in interpreting our results. As illustrated in Figure \ref{tree}, we allow any merger history between SN$_{prog}$ and SN$_{des}=63$.}



\section{Multi-merger fraction of the local Universe}
\label{mm_fraction}


\begin{table*}
\begin{minipage}{126mm}
	\caption[Table 2]{ Summary of merger fractions for the Millennium Run and the SDSS fraction estimated in this paper. The merger fraction depends on the snapshot number (SN$_{prog}$) of the progenitors which are recorded at redshifts ($z$) corresponding to look-back times ($\Delta t$) as given. To contribute to the numerator of a fraction, a galaxy must be volume-limited ($M_r<-20.55$) and be part of a major merger ($M_2/M_1<3$, $M_3/M_2<3$, $M_4/M_3<3$, etc.). *This estimate is based on too small a sample to reach any firm conclusions.	}	
		\label{table_2}
	\begin{tabular}{|| c | c | c | c | c | c | c | c | c | c || }
\hline \\
SN$_{prog}$ 	& $z$	& $\Delta t$ 	& Binary/Single & Triple/Binary	& Quad/Triple  		\\
			&   		& (Gyr)		& ($\%$) 		 & ($\%$) 		& ($\%$)  			\\
\hline \\
\multicolumn{6}{ l }{delucia06} 	\\
 58	& 0.116	& 1.0 	& 4.0 	& 7.1 	& 14.5	\\
 59	& 0.089 	& 0.8		& 3.1 	& 5.4 	& 12.5	\\
 60	& 0.064 	& 0.6		& \underline{2.2} 	& \underline{4.1} 	& \underline{7.8} 	\\
 61	& 0.041 	& 0.4		& 1.4 	& 3.0 	& 3.8 	\\
 62	& 0.020 	& 0.2		& 0.7 	& 1.5 	& 0		\\

\hline \\
\multicolumn{6}{ l }{bertone07} 	\\
 58	& 0.116 	& 1.0 	& 3.0		& 6.9 	& 14.2	\\
 59	& 0.089  	& 0.8		& 2.3		& 5.5		& 11.4 	\\
 60	& 0.064	& 0.6 	& \underline{1.6} 	& \underline{4.0} 	& \underline{7.8} 	\\
 61	& 0.041 	& 0.4		& 1.0	 	& 3.0 	& 3.4 	\\
 62	& 0.020 	& 0.2		& 0.5 	& 1.3 	& 0 	 	\\

\hline \\
\multicolumn{6}{ l }{bower06} 			\\
 58	& 0.116 	& 1.0		& 3.4		& 11.6	& 26.8	\\
 59	& 0.089 	& 0.8		& 2.6		& 8.8 	& 25.9	\\
 60	& 0.064 	& 0.6		& \underline{1.8} 	& \underline{7.4} 	& \underline{16.5} 	\\
 61	& 0.041 	& 0.4		& 1.1 	& 4.9 	& 10.4	\\
 62	& 0.020 	& 0.2		& 0.6 	& 2.7 	& 0		\\

\hline \\
\multicolumn{5}{ l }{D10a} 			\\
 -	&  - 	& - & $1.5-4.5 $		& $\lesssim 2 $	& $ 0 $ ($0 - 25$)*	 \\

\end{tabular}
\end{minipage}
\end{table*}


\subsection{SDSS Multi-Merger Fraction}
\label{mm_fraction_sdss}

In D10a, it was shown how the data from Galaxy Zoo can be used to measure the merger fraction of the local Universe by effectively integrating over the distribution of $f_m$. Galaxy Zoo data is only available for spectral targets and, since a meaningful merger-fraction needs to be volume-limited, only those galaxies which had acquired spectra were considered in the calculation {\color{black}(not all galaxies that are targeted for spectra obtain them due to fibre clashes in the SDSS apparatus).} This approach requires special consideration to the fact that only about $\sim 30 \%$ of the SDSS sky has spectroscopic completeness in so-called tile overlap regions ({\color{black}a section of the sky that has been viewed more than once in taking spectra}; see \citealt{strauss}; \citealt{blanton2}; D10a). In single-tile regions, the widths of the spectral fibres on the SDSS instrument do not permit spectral targets with in $55''$ of each other to both acquire spectra. One can correct for this effect with a multiplicative factor $C \sim 1.5$ (see Appendix A of D10a) to give a major merger fraction for the local Universe of $1 - 3 \times C \%$ for galaxies with $M_r<-20.55$.

We can estimate an upper limit for the multi-merger fraction simply by finding the ratio of volume-limited galaxies in major multi-mergers to volume-limited galaxies in binary-mergers in GZM1. This catalogue was constructed with $f_m>0.4$ and it is plausible to assume that multi-mergers are amongst the types of merger {\it more} likely to be classified as `merging' (in the Galaxy Zoo interface) than simple binaries because multi-mergers generally appear quite dramatic, prompting the user to go for the merger button. The fraction of multi-mergers in systems with $f_m>0.4$ is therefore likely to be greater than the fraction of multi-mergers in galaxies for all $f_m$. Therefore, by only considering $f_m>0.4$, we can estimate the upper limit of the multi-merger fraction in the nearby Universe.

When we volume limit the $2 \times 3003$ galaxies in GZM1 by the constraint $M_r < -20.55$, we are left with 1634 individual galaxies in major mergers (binary or multi). Of the 39 multi-mergers we have identified, only 16 are major triple-mergers and of these systems, $38/48$ have $M_r < -20.55$. This gives a fraction of $38/1634$ and so we can approximate the upper limit of the major triple-merger fraction as $ \lesssim 2 \%$.\footnote{More formally: the ratio of volume-limited ($M_r<-20.55$) galaxies in major triple-mergers to volume-limited galaxies in major binary-mergers is $\lesssim 2 \%$.} This number might be inflated by no more than $\sim 50 \%$ if one takes into account the few systems from the original 78 (see \S \ref{method}) that {\it might} have been multi-mergers, but could not be resolved sufficiently to be sure. The multi-merger to binary-merger ratio appears to be similar, therefore, to the binary-merger to single-galaxy ratio calculated in D10a ($1.5 - 4.5 \%$). We stress that this is a rough estimate for the upper boundary, since we have lost information by only considering systems for $f_m > 0.4$; but the general result is that the probability of finding a galaxy in a merger of $N$ galaxies (for $N=2, 3$) is $\sim$ few percent of the probability of finding a galaxy in a system of $N-1$ galaxies (for these volume-limited conditions $M_r < -20.55$ and $z<0.1$). 

Extending this empirical query to systems with $N \ge 4$ galaxies suffers from small number statistics. Of the 39 systems, we visually identify only 7 systems as having 4 or more galaxies merging at once. But none of these are {\it major}quadruple-mergers by the appropriate definition: $\mbox{M}_1/\mbox{M}_2<3$, $\mbox{M}_2/\mbox{M}_3<3$ and $\mbox{M}_3/\mbox{M}_4<3$. However, the systems on lines 3 and 19 of Table \ref{table1} only just miss out on this definition by having values of $\mbox{M}_3/\mbox{M}_4$ slightly above $3$ (with errors making $<3$ possible). If both of these systems {\it were} to be considered as major quadruple-mergers, then the ratio of volume-limited galaxies in major quadruple-mergers to galaxies in major triple-mergers would be $7\pm \sqrt{7} / 38$ (with $\pm \sqrt{7}$ being the poisson counting error). So while the {\it measured} quadruple-merger fraction is technically zero, it could easily have been as high as $\sim 25 \%$. Our sample is therefore too small to give accurate merger-fraction estimates for systems with 4 or more galaxies of comparable size.

\subsection{The Millennium Multi-Merger Fraction}

The galaxy databases for the Millennium Run offer much larger samples that allow us to calculate merger fractions out to quadruple-major mergers. For example, in the model of delucia06, at SN$_{desc}=63$ ($z=0$), there are 1,673,590 galaxies  with $M_r<-20.55$. The combined number of progenitor galaxies at SN$_{prog}=60$ is 1,943,561. As explained above, we classify each descendant galaxy at SN$_{desc}=63$ as a `merger remnant' if it has two or more progenitors at a given SN$_{prog}$ whose redshift corresponds to a look-back time comparable to a merger time-scale (we in fact vary SN$_{prog}$ between $58 - 62$). 

Of all these `mergers,' we further classify them as being `major' if all their progenitor masses are constrained by our working definition:  $\mbox{M}_2/\mbox{M}_1<3$, $\mbox{M}_3/\mbox{M}_2<3$, etc. It is clear that our Millennium merger-fractions will depend on the choice of SN$_{prog}$ since, the lower SN$_{prog}$ is, the greater the number of systems there are in mergers (as the merger tree branches out with increasing redshift). We therefore calculate the merger fractions for a range of SN$_{prog}$ values and present them in Table \ref{table_2} referring to fractions derived from SN$_{prog}=60$ (with look-back time of $\sim 0.6$ Gyr) as the fiducial value for each model. 

For the binary-single fraction, we find that all three SAMs produce a merger-fraction within the limits of D10a ($1.5-4.5 \%$). This is in agreement with similar comparisons to close-pairs (\citealt{kitz}; \citealt{patton2008}; \citealt{mateus}) and CAS (\citealt{bertone2}) for the local Universe. However, the SAMs give slightly higher percentages for the ratio of galaxies in triple-merger systems compared to binary-mergers. We estimated that this number should be no more than $\sim 2 \%$ but the SAMs have at least double that fraction for SN$_{prog}=60$. Only if we use the first Millennium time-step SN$_{prog}=62$, reducing the merger-detectability time-scale to $\sim 0.2$ Gyr do we get agreement between the MPA models and our calculation. This might be reconciled by the fact that multi-mergers are dominated by elliptical galaxies which have shorter merger-time scales (see \S \ref{properties}). The Durham model predicts roughly twice the number of multi-mergers than do the Munich models, so the latter are closer to our observations on multi-merger fractions. 

For the quadruple-triple ratio the Galaxy Zoo sample is too small to offer useful constraints. Comparing the Durham to MPA models shows, however, that the prior has a greater multi-merger fraction still (over $100 \%$ more). 



\begin{figure}
	\includegraphics[width=90mm]{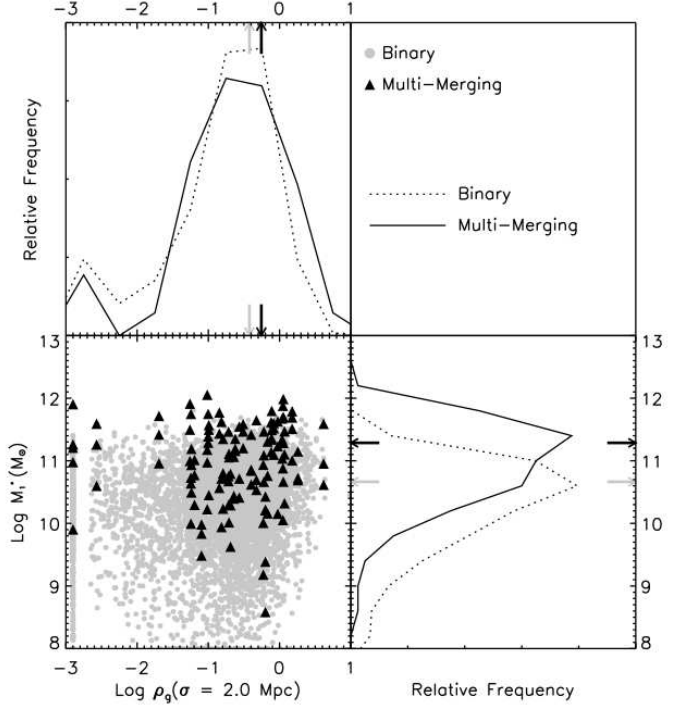}
	\caption[Figure 4]{Distributions of stellar mass and environment $\rho_g$ for galaxies in multi-mergers and galaxies in binary-mergers. The arrows point at the mean values for the distributions. The peak at $\log \rho_g = -3$ is artificial (to avoid $\log 0$ errors we set $\rho_g $ to $10^{-3}$ if zero).}
	\label{fig_mass_rho}
\end{figure}

\section{Properties of Multi-Merging Galaxies}
\label{properties}

\subsection{SDSS Multi-Merger Properties}
\label{sdss_properties}

\subsubsection{SDSS Morphologies}
\label{sdss_morph}

\begin{figure}
	\includegraphics[width=84mm]{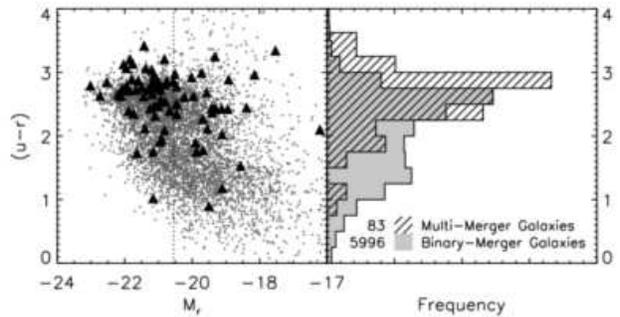}
	\caption[Figure 5]{Members of multimergers (triangles) overlaying the catalogue of binary-mergers (grey).}
	\label{fig_colour}
\end{figure}

One of four morphological categories was assigned to each galaxy in the multi-merger systems: S = spiral, E = elliptical, SU = `unsure' spiral and EU = `unsure' elliptical. Unsure morphologies are common, especially in late stage, distant systems where structural indicators like spiral arms cannot be seen. The same four categories were used in D10a where it was found that spirals (S and SU) outnumber ellipticals (E and EU) in volume limited galaxies in mergers by at least 3:1. This is about twice the ratio of the global population and it was argued in D10b that this is most likely due to the fact that mergers involving spirals remain detectable for longer periods than mergers involving ellipticals. In multi-mergers, by contrast, we found that of the spiral to elliptical ratio for volume-limited galaxies was closer to 1:1. In other words, multi-mergers appear to favour ellipticals in our sample compared with their occurrence in binary-mergers and the global galaxy population.


\subsubsection{SDSS Environment, Colours and Stellar Masses}
\label{sdss_enviro}

It seems unlikely that this high occurrence of ellipticals in multi-mergers is entirely due to a selection effect. It is possible that bulge dominated galaxies can be identified in multi-mergers longer than spirals since the bulge is a key feature indicating how many galaxies were originally involved in a multi-merger. Spirals remain detectable in mergers for longer (D10b) so the duration of detectability for multi-mergers involving spirals should be longer still than for multi-mergers involving ellipticals. Despite this, the fact that ellipticals feature so prominently in multi-mergers suggests that environment has some influence, i.e. dense environments are more likely to host multi-mergers. It was found in D10b that (binary) mergers tend to occupy slightly denser environments than galaxies in the global population and so it seems that merger number scales with environment.

We measure the environment of the multi-mergers directly using the {\it adaptive Gaussian environment parameter}, $\rho_{g}$, following D10b. This is a sophisticated measure of galaxy number density that utilises all the relevant spectral information in the SDSS database (\citealt{schawinski3}; D10b) returning a value $\rho_{g} \in \mathbb{R}^{+}$ for any coordinate in the SDSS sky ($ra, dec, z$). {\color{black}The parameter $\rho_g (ra, dec, z, \sigma)$ starts by finding close neighbours within an initial radius of $\sigma$ for each galaxy (we use $\sigma = 2.0$Mpc following \citealt{schawinski3}). It then adapts this radius depending on the initial return in order to compensate for the ``finger- of-God'' effect and is weighted such that $\rho_g$ increases the nearer its neighbours are. If $\rho_g=0$ then there are no neighbours within a $\sigma$ radius. A galaxy with $\rho_g = 1$ roughly corresponds to it being at the centre of a sphere of radius 3 Mpc with ten galaxies randomly distributed within (\citealt{schawinski3}).} 

Figure \ref{fig_mass_rho} shows the distribution of $\rho_{g}$ for the multi-merger and binary-merger systems (top-left panel). The multi-mergers on average occupy environments with slightly higher values of $\rho_g$. The Kolmogorov-Smirnov statistic between the binary and multi-merger data sets provides a measure of the difference between their cumulative distributions and we find them to be different with a significance level of $>99 \%$. 

The observed multi-mergers are therefore found in more dense environments though the primary reason is not clear: do multi-mergers favour ellipticals because they are likely to take place in dense environments or are multi-mergers with ellipticals easier to {\it visually identify} and therefore make multi-mergers {\it appear} to favour dense environments? Probably both factors are at play and it is difficult at this stage to disentangle the effects quantitatively.

Likewise, Figure \ref{fig_mass_rho} indicates that the stellar masses of the galaxies in multi-mergers are on average greater than their binary-merger counterparts. We compare the SDSS masses directly to the Millennium SAMs in \S \ref{mill_enviro}. Colour is yet another quantity that correlates with morphology and, as Figure \ref{fig_colour} shows, the $u-r$ colours of multi-merging galaxies are redder than their binary-merger counterparts. The empirical evidence is therefore emphatically clear that galaxies observed in multi-mergers are more likely to be early-type than single and binary-merger galaxies.

\begin{figure}
	\includegraphics[width=90mm]{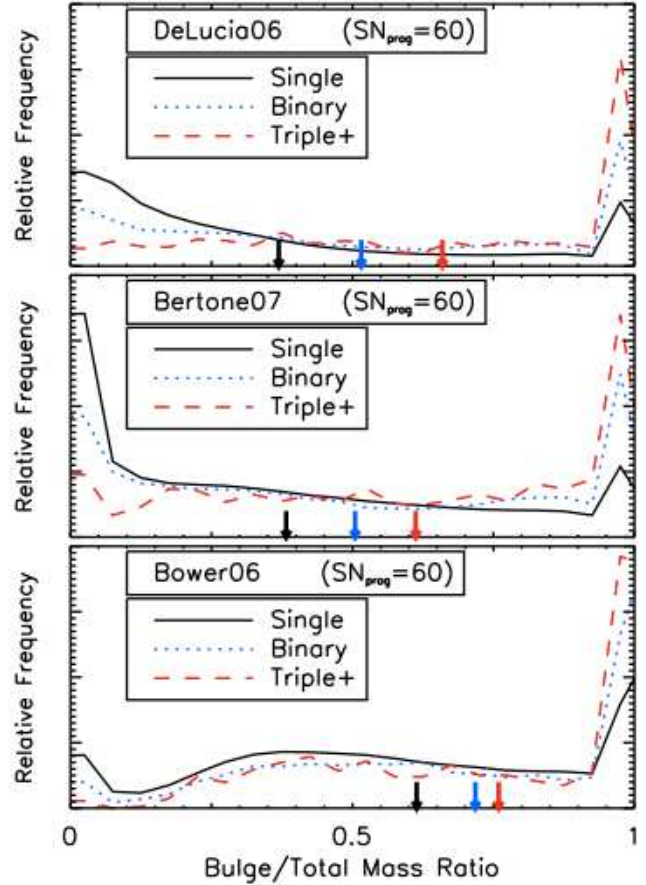}
	\caption[Figure 6]{Distributions of Bulge-Total mass ratio of galaxies in single-, binary- and multi-merger systems for all three SAMs at SN$_{prog}=60$. The arrows indicate the mean values for each sample. All three predict that binary and multi-mergers have more `elliptical' like morphologies compared with isolated systems - a likely concomitant of the favourability of mergers to occur in high-density environments.}
	\label{bulge_dist}
\end{figure}

\subsection{Millennium Multi-Merger Properties}
\label{mill_properties}

\subsubsection{Millennium Morphologies}
\label{mill_morph}

The morphologies of the SDSS multi-mergers and the Millennium multi-mergers cannot be compared directly since the SDSS morphologies are obtained visually. However, we can determine the qualitative relationship of the Millennium galaxy morphologies using the Bulge-Total stellar mass ratio as a proxy (with a high ratio corresponding to ellipticals and low ratio to spirals). Several studies have defined morphologies this way (\citealt{khochfar}; \citealt{benson3}; \citealt{bertone1}; \citealt{parry}). \citealt{benson3} found that the SDSS and Durham SAM produced qualitatively similar disc-bulge Luminosity Functions (see their Figure 17). The distributions  of Bulge/Total mass for the galaxies of the three SAMs at SN$_{prog}=60$ is shown in Figure \ref{bulge_dist}. All three models produce the expected qualitative result that the more galaxies there are in a (merger) system, the more bulge dominated they will be. The Durham model has systems generally more bulge-dominated in agreement with \citealt{parry}.

As argued in \S \ref{sdss_enviro}, binary- and multi-mergers occur most favourably in high-density environments and this is where interactions (inducing gravitational torques, see \citealt{hopkins}; \citealt{hopkins2}) causing disk instability are common. This is no doubt largely responsible for the well established morphology-environment relationship (established since at least \citealt{dressler}) and so mergers, taking place more favourably in denser environments, are more likely to be elliptical. 

\begin{figure*}
	\includegraphics[width=170mm]{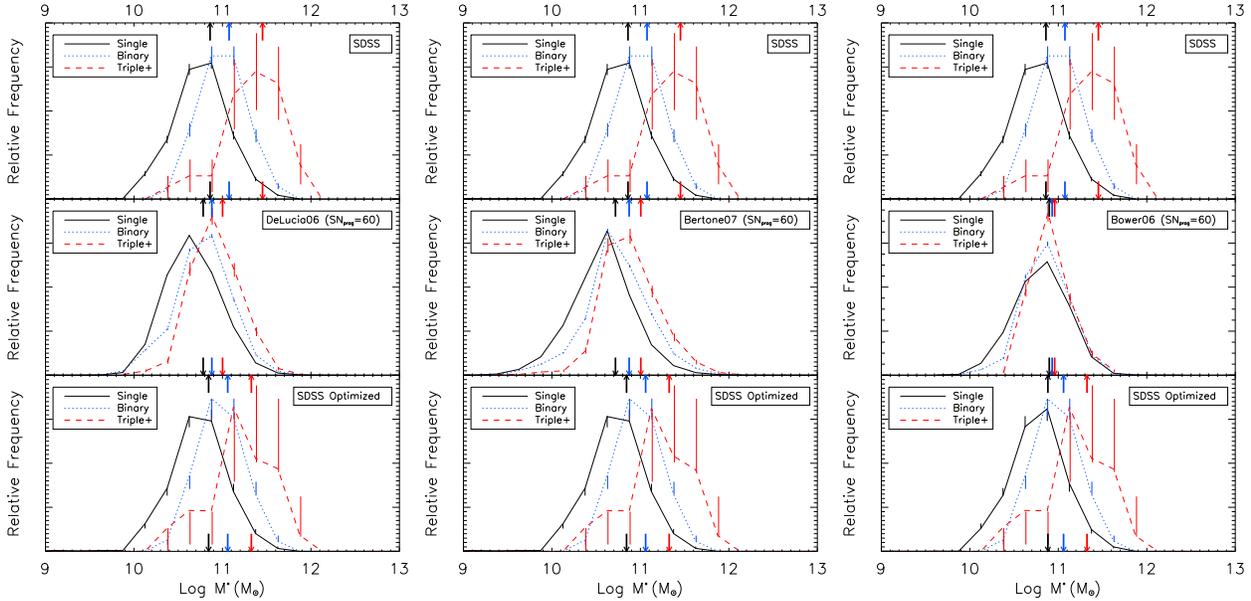}
	\caption[Figure 7]{Mass distributions for the single-, binary- and multi-merger systems in SDSS and Millennium. The left, middle and right columns correspond to the delucia06, bertone07 and bower06 models respectively. The top panels show the mass distributions for the SDSS systems. The bottom panels show the same (SDSS) data except the plotted points are allowed to vary along the error bars (representing the poisson counting errors) in order to minimise a {\color{black}$\chi ^2$} significance test with the Millennium model (in the middle panel). The arrows show the mean values for each distribution. All galaxies are volume-limited ($z<0.1, M_r<-20.55$) and the progenitors are taken at SN$_{prog}=60$. The Millennium SAM masses are determined here using $h^{-1}=1/0.73$.}
	\label{multi}
\end{figure*}

\subsubsection{Millennium Environment, Colour and Stellar Masses}
\label{mill_enviro}

The Millennium SAM catalogues do not provide a direct measure of environment and so we decline to test this property, though it seems almost certain that multi-mergers will favour high-density environments given the bulge-dominated morphologies exhibited by the SAMs as discussed in \S \ref{mill_morph}. 

The colours of delucia06 and bertone07 were examined in \citealt{bertone1} and it was found that bertone07 did not reproduce the colour bi-modality of the local Universe as well as delucia06. The model of bower06 does reproduce the local bi-modality well (see e.g. Figure 4 \citealt{bower2}). However, since the colours are derived secondarily from the stellar populations comprising their putative galaxies (all using the stellar synthesis models of \citealt{bruzual}), we analyse their stellar mass distributions as a main form of comparison.

Figure \ref{multi} shows the mass distributions for the SDSS and Millennium galaxies. \citealt{bertone1} reported that both their model and that of delucia06 slightly underestimate the Luminosity Function as a function of stellar mass in the range $10.6 \lesssim \log{\mbox{M}^*} \lesssim 11.2$. This helps explain the very slight discrepancy between the single-galaxy mass distributions for the MPA models (note the black vertical arrows show the mean masses of the SAMs are $\sim 1$ dex less than that of the SDSS single systems). The top and bottom panels show the SDSS distributions where the data points on the bottom are allowed to vary along the poisson-count error bars in order to most closely match the Millennium distributions. Also, importantly, the distributions for the binary-mergers, using the catalogue of D10a, are adjusted so as to include a spiral-elliptical ratio of $3:2$ since it was argued in D10b that spirals are over-observed by factor $\sim 2$ due to longer time-scales of detectability (see also \citealt{lotz3}).

By contrast, it was suggested in \S \ref{sdss_morph} that ellipticals might be over-observed but since we cannot quantify this, we cannot correct for it. It suggests that the multi-merger mass distributions might not increase at quite the rate suggested by the SDSS distributions. All this considered, the MPA models both do well in reproducing the fact (with slight underestimation) that galaxies in binary and multi-mergers increase in mass by $\sim 1$ dex per extra galaxy in the system (on average). By contrast, the model of bower06 appears to reproduce the single-galaxy mass distribution to great accuracy but shows very little in the way of increasing mass with merging.


\section{Summary}
\label{summary}

Through the Galaxy Zoo project, we have assembled a catalogue of multi-merger systems in the local Universe ($z<0.1$). Multi-mergers can only be found in SDSS through visual inspection (currently)  since close-pairs suffer from fiber overlaps and automated methods like CAS and GM$_{20}$ are not sensitive enough to consign a given image to a `multi-merger' parameter space (distinct from binary-mergers). The original Galaxy Zoo interface was not setup specifically for the task of finding multi-mergers, but will be in the future. This should produce a larger, more-complete catalogue.

Nonetheless, we argued that our catalogue is sufficient to provide a rough estimate of the (major) multi-merger fraction and gave an upper bound such that $\lesssim 2 \%$ of all volume-limited galaxies ($M_r<-20.55$) in a major merger (most of them being binary) are specifically in a {\it multi}-merger. This is about the same percentage as that calculated for the (binary) major merger fraction ($1.5-4.5\%$) for the same volume-limiting constraints (D10a). However, our sample is not large enough to find the quadruple-major merger fraction (or beyond).

The Millennium SAMs gave similar merger fractions for the number of volume-limited galaxies in major binary-mergers compared to single galaxies ($\sim 2 \%$). However, the next level of merger exhibited some disagreement between the SAMs and observation with the SAMs over-predicting galaxies in multi-mergers by at least factor $\sim 2$. The Durham model offered a multi-merger fraction twice that of the MPA model. However, since we have shown that galaxies in multi-mergers tend to be elliptical, which have shorter time-scales of detectability, this could justify taking the Millennium multi-merger step at SN$_{prog}=62$ in which case the fraction would be within our rough observational limit in the bertone07 model. 

Examining the properties of the galaxies in these multi-mergers to those in binary-mergers and single galaxy systems, we found that the (volume-limited) galaxies in multi-mergers have greater stellar masses (Figure \ref{multi}), redder colours (Figure \ref{fig_colour}) and occupied denser environments (Figure \ref{fig_mass_rho}). Such properties are characteristic of elliptical galaxies and we found a high proportion of ellipticals-spirals (at about $1:1$) in multi-mergers compared to the single systems (at about $2:3$) and binary-mergers (at about $1:3$). We argued that this is unlikely to be entirely due to a selection effect. To corroborate this, we compared our results with the SAMs. 

We found good qualitative agreement: all three models predicted that the bulge-total mass ratio increases with merger-number corroborating the observation that multi-mergers favour ellipticals and, implicitly, occupy denser environments on average compared to binary-mergers and single galaxies. The MPA models also agreed qualitatively with the fact that galaxy mass increases, on average, with the number of galaxies in the system. The Durham model appeared to slightly underestimate this effect.  

Future work with the Galaxy Zoo project aims to expand and improve these multi-merger catalogues as they provide a valuable, independent test to semi-analytical models.


\section{Acknowledgments}
\label{ack}

DWD acknowledges funding from the John Templeton Foundation. Kaviraj acknowledges a Research Fellowship from the Royal Commission for the Exhibition of 1851, an Imperial College Research Fellowship and a Senior Research Fellowship from Worcester College Oxford and a Leverhulme Early-Career Fellowship. CJL acknowledges support from the STFC Science in Society Program and the Leverhulme Trust. Support for the work of KS was provided by NASA through Einstein Postdoctoral Fellowship grant number PF9-00069 issued by the Chandra X-ray Observatory Center, which is operated by the Smithsonian Astrophysical Observatory for and on behalf of NASA under contract NAS8-03060.

Funding for the SDSS and SDSS-II has been provided
by the Alfred P. Sloan Foundation, the Participating Institutions,
the National Science Foundation, the U.S. Department
of Energy, the National Aeronautics and Space
Administration, the Japanese Monbukagakusho, the Max
Planck Society, and the Higher Education Funding Council
for England. The SDSS Web Site is http://www.sdss.org/.
The SDSS is managed by the Astrophysical Research Consortium for the Participating Institutions. The Participating
Institutions are the American Museum of Natural
History, Astrophysical Institute Potsdam, University of
Basel, University of Cambridge, Case Western Reserve University,
University of Chicago, Drexel University, Fermilab,
the Institute for Advanced Study, the Japan Participation
Group, Johns Hopkins University, the Joint Institute for
Nuclear Astrophysics, the Kavli Institute for Particle Astrophysics
and Cosmology, the Korean Scientist Group, the
Chinese Academy of Sciences (LAMOST), Los Alamos National
Laboratory, the Max-Planck-Institute for Astronomy
(MPIA), the Max-Planck-Institute for Astrophysics (MPA),
New Mexico State University, Ohio State University, University
of Pittsburgh, University of Portsmouth, Princeton
University, the United States Naval Observatory, and the
University of Washington.



\end{document}